\documentclass[journal]{IEEEtran}
\usepackage{amsmath}
\usepackage{amsfonts}
\usepackage{array}
\usepackage[caption=false,font=normalsize,labelfont=sf,textfont=sf]{subfig}
\usepackage{textcomp}
\usepackage{stfloats}
\usepackage{url}
\usepackage{verbatim}
\usepackage{graphicx}
\usepackage{cite}
\usepackage{xcolor}
\usepackage{amssymb}
\usepackage{bm}
\usepackage{makecell}
\usepackage{enumerate}
\usepackage{diagbox}
\usepackage{bm,bm,bbm}
\usepackage[colorlinks=true, linkcolor=blue, citecolor=blue, urlcolor=blue]{hyperref}
\usepackage{url}
\usepackage[outdir=./]{epstopdf}
\usepackage{fancyhdr}
\usepackage{mdwmath}
\usepackage{mdwtab}
\usepackage{amsthm}

\usepackage{cleveref}
\usepackage{booktabs}
\usepackage{tablefootnote}
\usepackage{pifont}
\usepackage{xspace}
\usepackage{subcaption}
\usepackage{setspace} 
\usepackage{balance}
\usepackage{multicol}
\usepackage{cuted}
\usepackage{mathtools}
\usepackage{float}
\usepackage{multirow}
\usepackage[font=small]{caption} 
\usepackage{algorithm}
\usepackage{algpseudocode}
\usepackage{enumitem}
\usepackage{orcidlink}
\usepackage{threeparttable}
\usepackage{textcase}

\begin{document}

\title{GLo-MAPPO: Multi-Agent Deep Reinforcement Learning for Energy-Efficient UAV-Assisted LoRa Networks}

\author{Abdullahi~Isa~Ahmed,~\IEEEmembership{Graduate~Student~Member,~IEEE,}
Jamal~Bentahar,~\IEEEmembership{~Member,~IEEE} and~El~Mehdi~Amhoud,~\IEEEmembership{Senior~Member,~IEEE}

\thanks{Copyright (c) 2026 IEEE. Personal use of this material is permitted. However, permission to use this material for any other purposes must be obtained from the IEEE by sending a request to pubs-permissions@ieee.org.}

\thanks{Abdullahi Isa Ahmed and El Mehdi Amhoud are with the College of Computing, Mohammed VI Polytechnic University, Ben Guerir, Morocco. (e-mail: {\tt\small abdullahi.isaahmed@um6p.ma, elmehdi.amhoud@um6p.ma}).}

\thanks{Jamal Bentahar is with Advanced Intelligent Systems Research Center (KU-AIS), Department of Computer Science, Khalifa University, UAE, and also with Concordia Institute for Information Systems Engineering, Concordia University, Montreal, QC, Canada (email: {\tt\small jamal.bentahar@ku.ac.ae})}}

\markboth{%
  \protect\NoCaseChange{
    \makebox[\textwidth][c]{
      \normalfont\small
      \shortstack{
        \textcopyright~2026 IEEE. This article has been accepted for publication in IEEE Transactions on Vehicular Technology in October 2026.
      }
    }
  }
}{}

\maketitle

\begin{abstract}
The rapid advancement of Low-Power Wide Area Networks (LPWANs), particularly Long Range (LoRa) technology, has positioned them as a cornerstone of Next-Generation Internet of Things (NG-IoT) applications. Despite their long-range and low-power advantages, achieving high energy efficiency (EE) in LoRa networks remains a significant challenge in highly dynamic environments. Traditional terrestrial gateway deployments often suffer from coverage gaps and non-line-of-sight (NLoS) propagation, while satellite-based alternatives incur excessive energy consumption and prohibitive latency. To address these limitations, we propose a multi-UAV architecture in which unmanned aerial vehicles (UAVs) serve as mobile LoRa gateways to dynamically collect data from LoRa end devices (EDs). We formulate a joint optimization problem to maximize the weighted system EE by jointly optimizing spreading factors, transmission powers, UAV trajectories, and ED-UAV associations. This problem is transformed into a partially observable stochastic game (POSG), which we solve using our proposed Green LoRa Multi-Agent Proximal Policy Optimization (GLo-MAPPO). We adopt PPO because its clipped surrogate objective discourages excessively large policy updates, thereby promoting stable learning as multiple UAV agents simultaneously adapt their behaviors. Combined with centralized training and decentralized execution (CTDE), this enables coordinated policy learning while allowing each UAV to make decisions independently during execution. Simulation results show that GLo-MAPPO consistently outperforms state-of-the-art multi-agent reinforcement learning (MARL) benchmarks in terms of EE and power consumption across varying network densities. We further validate the trained policy through closed-loop, packet-level co-simulation using the OMNeT++/FLoRa LoRa simulator. Finally, an ablation study highlights the contribution of each optimization component and demonstrates the effectiveness of the proposed association scheme.
\end{abstract}


\section{Introduction} \label{Section_1} 
\IEEEPARstart{T}{he} Internet of Things (IoT) has transformed connectivity across many sectors, enabling applications in smart systems and beyond \cite{11037543}. With the number of connected IoT devices projected to reach roughly $125$ billion by 2030 \cite{fakhruldeen2024enhancing}, global energy consumption is rising rapidly. Developing energy-efficient IoT communication frameworks has therefore become a critical priority, directly supporting United Nations Sustainable Development Goal 7 (UN SDG 7) on affordable and clean energy \cite{orazi2018first}.

In response to this, Low-Power Wide Area Networks (LPWANs) have emerged as key enablers for sustainable IoT, offering wide coverage at low power \cite{jouhari2023survey}. The LPWAN landscape splits into two paradigms by spectrum use. Licensed technologies, such as Narrowband IoT (NB-IoT) and Long-Term Evolution for Machines (LTE-M), reuse cellular infrastructure for high reliability, but at significant deployment and subscription cost. In contrast, unlicensed solutions like Long Range (LoRa) prioritize energy autonomy and low cost, making them well suited to large-scale green deployments. LoRa in particular uses the modulation of the chirp spread spectrum (CSS) and adaptive spreading factors to balance the communication range against the data rate \cite{10183362}.

Despite these advantages, traditional LoRa deployments rely on static terrestrial Gateways (GWs), which struggle in Non-Line-of-Sight (NLoS) and dynamic settings. Fixed infrastructure adapts poorly to changing propagation, which leads to inefficient resource use, lower reliability, and wasted energy. Satellite-based solutions offer global coverage, but demand higher transmission power and add significant latency, limiting their use for energy-constrained applications.

To overcome these limits, aerial GWs mounted on Unmanned Aerial Vehicles (UAVs) have drawn strong interest \cite{11106313}. UAVs can adapt their trajectories to avoid NLoS conditions and improve link quality \cite{zhang2018fast}. Beyond connectivity, UAVs are increasingly explored across next-generation wireless paradigms, including cooperative terahertz (THz) communication for 6G and beyond~\cite{mustari2024cooperative}, reflecting the wide range of bands and applications they are being adapted to. While UAVs are well studied for cellular coverage, their use in LPWAN, and specifically in LoRa networks, is still at an early stage. Most existing work considers static configurations or a single UAV, leaving multi-UAV cooperation largely unaddressed. The few studies that use multiple aerial LoRa gateways tend to overlook the tight coupling between trajectory and resource management, or ignore the non-stationarity of multi-agent coordination. As a result, a framework that combines cooperative mobility with LoRa-specific resource constraints is still missing.

Optimizing such decentralized systems requires resource management approaches capable of handling coupled decision variables and dynamically changing network conditions. For example, conventional methods such as alternating optimization (AO) generally solve the original problem through a sequence of simpler subproblems. However, their performance may depend on the initialization, and they typically converge to a locally optimal solution without guaranteeing global optimality. In contrast, deep reinforcement learning (DRL) provides a promising alternative for adaptive decision-making in high-dimensional and dynamic environments \cite{11175095}. By interacting with the environment, DRL can learn high-quality decision policies while substantially reducing the online computational burden after training. Furthermore, single-agent methods such as Deep Q-Networks (DQN) scale poorly to multi-UAV deployments. As several UAVs learn at once, the environment becomes non-stationary, the best strategy for one agent shifts as the others change their own. This coupling, together with the partial observability of the environment, makes single-agent methods prone to divergence. Multi-agent methods are therefore needed to achieve genuine cooperation.

To address these challenges, we propose Green LoRa Multi-Agent Proximal Policy Optimization (GLo-MAPPO), a multi-agent reinforcement learning (MARL) framework built on MAPPO. It jointly optimizes spreading factors, transmission powers, UAV trajectories, and End-Device (ED) associations. We model the interaction as a Partially Observable Stochastic Game (POSG) and solve it with a Centralized Training with Decentralized Execution (CTDE) scheme for stable convergence. To the best of our knowledge, this is the first dynamic multi-UAV LoRa framework to jointly optimize gateway mobility, physical-layer resource allocation, and device association in unlicensed bands. Our framework extends a MAPPO-based MARL approach by integrating a LoRa-specific problem formulation, device association strategy, reward design, and a closed-loop validation scheme into the learning framework. Our main contributions are as follows:

\begin{itemize} 
\item We formulate a multi-objective problem to maximize the system's weighted energy efficiency. Unlike prior work, our model captures the tight coupling between LoRa physical-layer parameters and network-layer dynamics under a realistic probabilistic ground-to-air (G2A) propagation model. To this end, we design a multi-component reward that jointly shapes coverage, energy efficiency, collision avoidance, and mission completion toward deployable, energy-feasible trajectories.

\item We transform the multi-UAV coordination problem as a POSG scheme to capture the uncertainty of dynamic IoT environments, and solve it with GLo-MAPPO, a MARL framework that uses CTDE for stable, cooperative decision-making under non-stationarity.

\item We validate the trained policy beyond the analytic training environment through a closed-loop, packet-level co-simulation with OMNeT++/FLoRa, coupled over sockets in a lock-step loop. Across three channel tiers of increasing realism, the co-simulation reproduces the standalone energy-efficiency, delivery, and coverage results to within a fraction of a percent, confirming that the learned policy transfers to a packet-level LoRa stack.

\item Through extensive simulations, we show that GLo-MAPPO achieves superior convergence, energy efficiency, and low power consumption against state-of-the-art MARL benchmarks. An ablation study further quantifies the energy efficiency gains from the trajectory, resource allocation, and association scheme.
\end{itemize}

The remainder of this paper is organized as follows. Section~\ref{Section_2} reviews related work. Sections~\ref{Section_3} and~\ref{Section_4} present the system model and problem formulation. Section~\ref{Section_5} gives the POSG formulation, and Section~\ref{Section_6} details the GLo-MAPPO algorithms. Finally, Sections~\ref{Section_7},~\ref{Section_8}, and~\ref{Section_9} present the simulation results, discussion and future work, and conclusions, respectively.

\section{Related Works} \label{Section_2}
In this section, we first review deployment strategies for LoRa GWs. Then, we further extend that discussion to joint optimization of communication parameters and mobility. Lastly, we discuss recent advances in intelligent techniques, particularly multi-agent frameworks, and identify their potential for efficient LoRa network optimization.

\subsection{Gateway Deployment in LoRa Networks}
The performance of LoRa networks depends heavily on how GWs are placed and operated. In the conventional LoRa deployments, GWs are usually fixed terrestrial stations installed at permanent locations. Although these setups are straightforward to implement and cost-effective, they often fail to deliver reliable coverage, particularly in rural, remote, or rapidly changing environments. In such scenarios, factors such as challenging terrain, physical obstacles, among others, can seriously weaken signal quality and reduce overall network performance \cite{abbasi2024lora}.

To overcome this limitation, researchers have investigated alternative solutions. One line of work turns to non-terrestrial, satellite-based LoRa deployments \cite{alvarez2022uplink, tondo2024multiple, 11165344}. These systems, however, draw high power, add latency, and require tight synchronization \cite{9537682}, which makes them impractical for many IoT use cases. To further enhance efficiency, improve line-of-sight (LoS) communication, and support flexible deployment in the next generation of IoT networks, UAV-mounted LoRa GWs have emerged as an alternative solution that offers flexible deployment, reduced delay, and minimized power consumption. However, most existing works focus on a single UAV operating in static hover mode \cite{marchese2019flying, XIONG2023109511, jouhari2023deep}, which leads to an underutilization of UAV mobility. For instance, the authors in \cite{marchese2019flying} propose integrating a single UAV with LoRaWAN technology to extend IoT network coverage in remote and hard-to-reach areas. In \cite{XIONG2023109511}, the authors propose an energy-efficient LoRa resource allocation and single UAV trajectory design to minimize energy consumption. Nevertheless, these studies often lack coordination among multiple UAVs.

To address these limitations, our work departs from single-UAV or static hovering approaches by considering a multi-UAV mobile framework. Unlike prior studies that utilize UAVs merely as static aerial relays, we leverage UAV mobility to dynamically serve ground end devices. This collaborative design specifically addresses the lack of coordination in existing literature by establishing a collaborative multi-gateway environment, ensuring reliable coverage in dynamic scenarios where single or static GW solutions are insufficient.
\begin{table*}
\scriptsize 
\centering
\caption{\small{Comparison of Our Proposed Work with Existing Approaches}}
\label{literature_review}
\begin{threeparttable}
\resizebox{\textwidth}{!}{ 
    \begin{tabular}{|c|c|c|c|c|c|c|c|c|c|c|} \hline  
        \multirow{2}{*}{\textbf{Work}} & \multirow{2}{*}{\textbf{Year}} & \multicolumn{4}{c|}{\textbf{Optimization Scope\textsuperscript{*}}} & \multirow{2}{*}{\textbf{\makecell{Mobile \\ Gateway}}} &  \multirow{2}{*}{\textbf{\makecell{Partial \\ Observation}}} & \multirow{2}{*}{\textbf{\makecell{Control \\ Approach}}} & \multirow{2}{*}{\textbf{Algorithm}} & \multirow{2}{*}{\textbf{Performance Metrics\textsuperscript{$\dagger$}}} \\ \cline{3-6}
        & & SF & TP & UAV Trj. & ED Asso. & & & & & \\ \hline  
        \cite{alvarez2022uplink} & 2022 & \ding{51} & \ding{51} & \ding{55} & \ding{55} & \ding{51} & \ding{55} & Centralized & Heuristic & Energy consumption, PER \\ \hline  
        \cite{marchese2019flying}& 2019 & \ding{55} & \ding{55} & \ding{55} & \ding{55} & \ding{55} & \ding{55} & Centralized & - & Quality of service \\ \hline  
        \cite{XIONG2023109511}& 2023 & \ding{51} & \ding{51} & \ding{51} & \ding{55} & \ding{51} & \ding{55} & Centralized & SCA+greedy & Energy consumption\\ \hline  
        \cite{jouhari2023deep}& 2023 & \ding{51} & \ding{51} & \ding{55} & \ding{55} & \ding{55} & \ding{55} & Centralized & PPO & Energy efficiency \\ \hline  
        \cite{amichi2019spreading}& 2019 & \ding{51} & \ding{55} & \ding{55} & \ding{55} & \ding{55} & \ding{55} & Centralized & Heuristic & Throughput, fairness \\ \hline  
        \cite{hamdi2020dynamic}& 2020 & \ding{51} & \ding{55} & \ding{55} & \ding{55} & \ding{55} & \ding{55} & Centralized & Greedy, CSI-based heuristic & Symbol error rate \\ \hline  
        \cite{su2020energy}& 2020 & \ding{51} & \ding{51} & \ding{55} & \ding{55} & \ding{55} & \ding{55} & Centralized & Matching theory+Heuristic & System energy efficiency \\ \hline  
        \cite{arasu2022efficient}& 2022 & \ding{51} & \ding{55} & \ding{55} & \ding{55} & \ding{55} & \ding{55} & Centralized & OMILP-LoRa & Energy consumption, throughput \\ \hline  
        \cite{9860807}& 2022 & \ding{51} & \ding{51} & \ding{55} & \ding{55} & \ding{55} & \ding{55} & Decentralized & MIX-MAB with SE & PDR\\ \hline  
        \cite{10540417}& 2024 & \ding{51} & \ding{51} & \ding{55} & \ding{55} & \ding{55} & \ding{55} & Centralized & IR-LEOA & Throughput, energy efficiency\\ \hline  
        \cite{aihara2019q} & 2019 & \ding{51} & \ding{55} & \ding{55} & \ding{55} & \ding{55} & \ding{55} & Centralized & Deep Q-learning & PDR \\ \hline  
        \cite{ZHAO2023100776}& 2023 & \ding{51} & \ding{51} & \ding{55} & \ding{55} & \ding{55} & \ding{55} & Decentralized & Multi-agent DQN & EPP and DER \\ \hline  
        \cite{10437219} & 2023 & \ding{51} & \ding{51} & \ding{55} & \ding{55} & \ding{55} & \ding{55} & Decentralized & MAAC & Energy efficiency, packet delivery rate \\ \hline 
        \cite{yu2020multi} & 2020 & \ding{51} & \ding{51} & \ding{55} & \ding{55} & \ding{55} & \ding{55} & Decentralized & Multi-agent Q-learning & Power consumption \\ \hline  
        \textcolor{black}{\textbf{Our work}} & \textcolor{black}{\textbf{2026}} & \textcolor{black}{\textbf{\ding{51}}} & \textcolor{black}{\textbf{\ding{51}}} & \textcolor{black}{\textbf{\ding{51}}} & \textcolor{black}{\textbf{\ding{51}}} & \textcolor{black}{\textbf{\ding{51}}} & \textcolor{black}{\textbf{\ding{51}}} & \textcolor{black}{\textbf{CTDE}} & \textcolor{black}{\textbf{GLo-MAPPO}} & \textcolor{black}{\textbf{System energy efficiency, power consumption, collision, PDR}}
        \\ \hline 
    \end{tabular}
}
\begin{tablenotes}
\footnotesize
\item[*] In the \emph{Optimization Scope} column, \emph{SF} is spreading factor, \emph{TP} is ED transmission power, \emph{UAV Trj.} is UAV trajectory and \emph{ED Asso.} is ED association.
\item[$\dagger$] In the \emph{Performance Metrics} column, 
\emph{PER} is packet error rate,
\emph{PDR} is packet delivery ratio
\emph{EPP} is energy per packet, 
\emph{DER} is data extraction rate.
\end{tablenotes}
\end{threeparttable}
\vspace{-0.5cm}
\end{table*}
\subsection{Joint Optimization of UAV Path and LoRa Parameters}
Another line of research explored how efficient resource allocation is essential for optimizing LoRa networks, particularly in dynamic or large-scale deployments. Many studies focus narrowly on optimizing only spreading factor, carrier frequency and spreading factor \cite{amichi2019spreading, hamdi2020dynamic}, or at best, jointly optimizing user scheduling, spreading factor, and transmission power \cite{su2020energy, arasu2022efficient}. More specifically, the authors in~\cite{su2020energy} proposed an energy-efficient uplink transmission framework that simultaneously optimizes user scheduling, spreading factor assignment, and power allocation through a low-complexity scheduling algorithm using matching theory. Besides, the work in ~\cite{arasu2022efficient} proposed a mixed-integer linear programming approach (OMILP-LoRa) with their ACCURATE-based heuristic framework to dynamically adjust spreading factor and carrier frequency configurations to minimize power consumption. However, these studies often treat GW positioning as either static or independent from communication parameter optimization. 

In the increasingly dynamic IoT environment, separating GW placement or trajectory planning from other optimizations is problematic, as the GW's location directly impacts link quality, path loss, and the energy consumption of end devices. Existing studies that overlook joint GW optimization often lead to suboptimal system performance. Moreover, this drawback can be attributed to inefficient optimization techniques that struggle with high-dimensional parameters. For instance, traditional methods, such as alternating optimization, decouple parameters into independent sub-problems to simplify the objective function \cite{bezdek2003convergence}. However, this approach frequently converges to suboptimal solutions and becomes computationally expensive as the number of variables increases. 

In contrast to these decoupled approaches from earlier studies, we jointly optimize LoRa physical layer parameters and UAV trajectories. Specifically, by treating these variables as interdependent rather than isolating them. Hence, our framework overcomes the local optima limitations inherent in conventional alternating optimization techniques. Beyond conventional and alternating-optimization techniques, learning-based trajectory generation has also been explored, such as visual generative adversarial networks for end-to-end UAV trajectory design in RIS-assisted energy-efficient vehicular networks~\cite{eskandari2025visual}. While effective for high-rate vehicular links, such approaches are not tailored to the spreading-factor and duty-cycle constraints that govern energy-efficient LoRa operation, which our framework addresses directly.

\subsection{Multi-UAV Mounted LoRa Networks with MARL}
Thanks to the emergence of Artificial Intelligence (AI), specifically DRL, which has been established to achieve optimal decision-making and maximize long-term rewards. In addition, DRL offers a powerful data-driven alternative, enabling systems to learn optimal policies through continuous interaction with the environment \cite{10628007}. Unlike conventional methods, DRL's ability to process high-dimensional spaces allows for a more joint optimization approach, capturing interdependencies between variables that are often lost when parameters are fully decoupled. Furthermore, recent works in DRL \cite{9860807, 10540417, aihara2019q, jouhari2023deep}, such as the one proposed by \cite{9860807}, develop an RL-based distributed resource allocation via MIX-MAB to boost packet delivery ratios. Similarly, the authors in \cite{10540417} leverage DQN for adaptive optimization of power, channel, and spreading factor to reduce energy consumption. Besides, the work in \cite{aihara2019q} apply Q-learning to reduce interference among nodes with rewards tied to successful packet reception. A proximal policy optimization (PPO) algorithm for dynamic spreading factor and transmission power allocation to enhance energy efficiency over air-to-ground links was proposed in \cite{jouhari2023deep}. However, single-agent DRL approaches mostly, have certain limitations in distributed settings. In particular, they often assume full observation of the environment, which is unrealistic in practical deployments. In fact, the reality is that most GWs operate under partial observability, where each agent only has access to local information. This restricts their ability to make globally optimal decisions and impairs coordination. 
\begin{table}[t]
    \centering
    \caption{Summary of Important Notations.}
    \label{notations_table}
    \footnotesize
    \begin{tabular}{|>{\centering\arraybackslash}m{1.6cm} |>{\raggedright\arraybackslash}m{6.2cm}|}
        \hline 
        \textbf{Symbol} & \textbf{Description} \\ \hline 
        $\mathcal{V}$, $V$, $v$ & Set, number, index of LoRa end devices (EDs)\\ \hline 
        $\mathcal{U}$, $U$, $u$ & Set, number, index of flying LoRa gateways (UAVs)\\ \hline
        $\mathcal{T}$, $T$, $t$ & Set, number, index of time-slots\\ \hline
        $H_u$, $f$ & Altitude of UAV $u$, carrier frequency\\ \hline 
        $\phi[t]$ & Elevation angle from UAV to ED at time $t$\\ \hline 
        $d_{u,v}[t]$ & Distance between UAV $u$ and ED $v$ at time $t$\\ \hline 
        $D_{\text{safe}}$ & Minimum safety distance between two UAVs\\  \hline
        $P^{\textrm{LoS}}_{u,v}$, $P^{\textrm{NLoS}}_{u,v}$ & LoS and NLoS probabilities between UAV and ED \\ \hline 
        $L^{\text{G2A}}_{u,v}[t]$ & G2A path loss for UAV $u$ and ED $v$ link at time $t$\\ \hline 
        $G_{u,v}[t]$ & Channel gain for UAV $u$ and ED $v$ link at time $t$\\ \hline 
        $a_{u,v}[t]$ & Association indicator between UAV and ED at time $t$\\ \hline 
        $p_{v,j}[t]$ & Transmit power of ED using power index $j$ at time $t$\\ \hline 
        $\psi_{v,n}[t]$ & Indicator that ED $v$ uses spreading factor $n$ at time $t$ \\ \hline  
        $\bar{\mathbf{\Psi}}$, $\bar{\mathbf{P}}$ & Finite sets of available spreading factors and power\\ \hline 
        $P_{\textrm{prop}}$, $P_{\textrm{max}}$ & UAV propulsion power consumption and maximum power budget \\ \hline
        $\rho_{u,v}[t]$ & SNR between UAV $u$ and ED $v$ at time $t$ \\ \hline
        $\bar{\Omega}_n$ & Per-SF reception (SNR) threshold \\ \hline
        $q^{\text{co}},\ q^{\text{in}}_n$ & Co-SF and inter-SF capture thresholds \\ \hline
        $T_{\text{air}}(n), T_{\text{dc}}, \eta_{\text{dc}}$ & Time-on-air, airtime budget, and duty cycle \\ \hline
        $\Re_{u,v}[t]$ & Achievable data rate for UAV $u$-ED $v$ link at time $t$ \\ \hline
        $EE_{\textrm{sys}}$ & System weighted energy efficiency\\ \hline 
    \end{tabular}
    \vspace{-10pt}
\end{table}
To address the limitations, a few studies have proposed MARL approaches for LoRa networks \cite{ZHAO2023100776, 10437219, yu2020multi}. For instance, the author in \cite{ZHAO2023100776} proposes a MARL algorithm to optimize energy efficiency in wireless underground sensor networks by jointly considering link quality, collisions, and energy consumption in dynamic underground environments. Similarly, the authors in \cite{10437219} introduces MALoRa, a MARL LoRa framework with an attention mechanism that enables end devices to focus on relevant peers' actions for better parameter allocation and system energy efficiency. Besides, the authors in \cite{yu2020multi} present a fully distributed multi-agent Q-learning scheme for dynamic power and spreading factor allocation to reduce collisions and improve reliability in dense deployments. 

Despite their promise, none of the existing works effectively handle the challenges of mobile gateways operating under partial observability. Unlike prior MARL works that rely on either fully decentralized learning or centralized execution, our approach adopts the CTDE paradigm. This strategy enables agents to utilize global information during training while acting on local partial observations during execution, thereby resolving the coordination bottlenecks of previous approaches. Motivated by these limitations, we propose our GLo-MAPPO framework, which jointly optimizes LoRa physical layer parameters, ED association, and UAV trajectories under POSG formulation to maximize system energy efficiency. Table \ref{literature_review} presents a comparative analysis between our proposed approach and existing methods. Although MAPPO has been applied to UAV communication tasks, such frameworks typically target cellular or generic wireless links and do not widely integrate it into the LoRa-specific physical-layer structure, nor do they jointly couple spreading-factor and transmission-power selection with trajectory and device-association decisions. In contrast, our framework integrates the LoRa PHY constraints directly into the POSG formulation and augments the MAPPO loop with a structured gain-based association layer, which distinguishes it from a direct application of MAPPO to a LoRa scenario.

\vspace{-2.0mm}
\section{System Model}\label{Section_3}

\begin{figure}[t!]
  \centering 
  \includegraphics[width=0.8\linewidth]{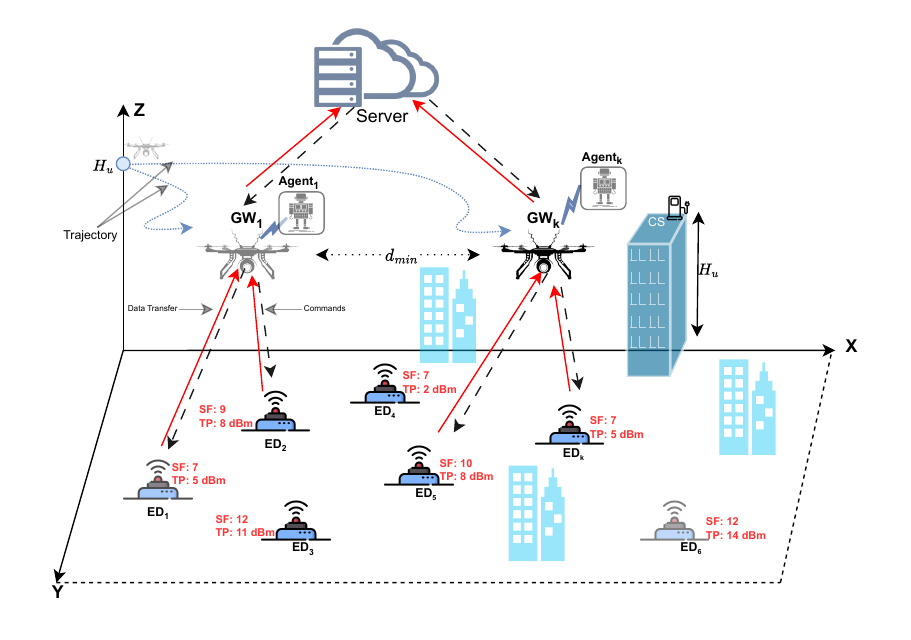} 
  \caption{The studied system model.}
  \label{fig_system_model}
  \vspace{-15pt}
\end{figure}

As shown in Fig.~\ref{fig_system_model}, we consider an uplink multi-UAV-mounted flying LoRa network consisting of $V$ LoRa EDs and $U$ flying GWs, which are defined as the set $\mathcal{V} \triangleq \{1, 2, \ldots , V  \}$ and the set of $ \mathcal{U} \triangleq \{1, 2, \ldots , U  \}$, respectively. We assume that the flying GWs are deployed in a designated target area to provide coverage to EDs randomly distributed within this region. Furthermore, the flying UAVs are assumed to be equipped with dual-function systems, which include a radar sensor capable of sensing the environment for collision avoidance and a wireless LoRa module for communicating with ground EDs. Additionally, a flying GW can serve multiple EDs within its communication range $R_{comm}$, collect data, decode packets, and then forward it to a centralized network server. Note that Table~\ref{notations_table} summarizes the most relevant notations.
\vspace{-5mm}
\subsection{Mobility Model} \label{mobility_model}
In this work, the EDs are assumed to be static and are generated randomly within the target area. On the other hand, flying GWs are mobile. Specifically, GWs begin from an initial position $\mathbf{q}_u^{\text{init}}$, collect data generated by the EDs\footnote{We assume that the data generated by each device is stored in a buffer until collected by a UAV.}, operate within a specified flight duration, and then return to a charging station (CS) $\mathbf{q}_{\text{cs}}$ to recharge for subsequent flight missions. Furthermore, we denote the total flight duration as $T_{\text{total}}$ and discretize it into $T$ equal time steps, each of duration $\Delta t = T_{\text{total}}/T$. Accordingly, the time steps belong to the set $\mathcal{T} = \{1, 2, \ldots, T\}$. Additionally, we adopt a 3D coordinate system to represent the locations of both flying GWs and EDs. The position vector of a GW $u$ at time slot $t \in \mathcal{T}$ is given by $\mathbf{q}_u[t] = [x_u[t], y_u[t], H_u]^T$, where $H_u$ denotes its fixed flying altitude. Conversely, the position of an ED $v$ is denoted by $\mathbf{g}_v = [x_v, y_v, 0]^T$, assuming it is located on the ground plane.

\begin{figure}[t!]
  \centering
  \includegraphics[width=0.6\linewidth]{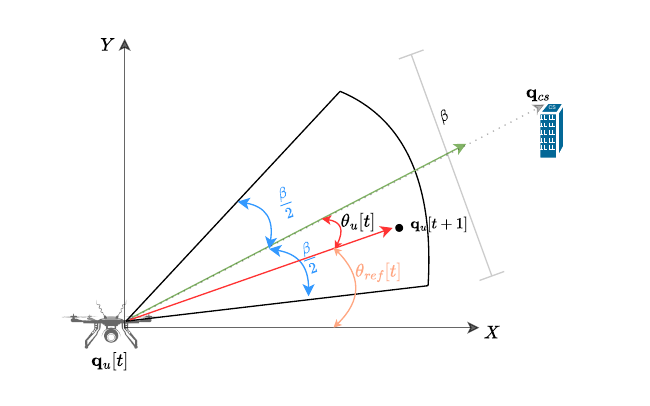} 
  \caption{UAV-assisted LoRa GW cone-constrained movement toward the charging station.}
  \label{fig_cone_movement}
  \vspace{-15pt}
\end{figure}

As illustrated in Fig.~\ref{fig_cone_movement}, at time $t$, a UAV located at position $\mathbf{q}_u[t]$ selects a direction $\theta_u[t]$ and moves at a speed $s_u[t] \in [0, S_{\max}]$ to reach its next position $\mathbf{q}_u[t+1]$. The movement direction $\theta_u[t]$ is defined as the angle between the vector from the UAV to the CS, i.e., $\mathbf{q}_{\text{cs}} - \mathbf{q}_u[t]$, and the displacement vector $\mathbf{q}_u[t+1] - \mathbf{q}_u[t]$. This direction is constrained within a cone, i.e., $\theta_u[t] \in [-\frac{\beta}{2}, \frac{\beta}{2}]$, where $\beta \in [0, 2\pi]$ is the cone's total angular width. Moreover, the reference angle $\theta_{\text{ref}}[t]$, i.e., the heading of the displacement vector with respect to the $X$-axis, is obtained by offsetting the bearing toward the CS by $\theta_u[t]$, expressed as $\theta_{\text{ref}}[t] = \operatorname{atan2}\!\left( y_{cs} - y_u[t],\; x_{cs} - x_u[t] \right) + \theta_u[t]$, where $\operatorname{atan2}(\cdot,\cdot)$ is the four-quadrant arctangent returning the CS bearing over $(-\pi,\pi]$. Note that the position of the $u$-th UAV at time slot $t$ is updated according to the following expression
\begin{equation}
\begin{cases}
x_u[t+1] = x_u[t] + s_u\cos(\theta_{\text{ref}}[t]) \Delta t, \\
y_u[t+1] = y_u[t] + s_u\sin(\theta_{\text{ref}}[t]) \Delta t. 
\end{cases}
\label{eq:position_update_uav}
\end{equation}

As the UAVs move, their positions evolve dynamically to optimize data collection from the EDs. To ensure feasible and safe UAV flight paths, several constraints are imposed. We define $\dot{\mathbf{q}}_u[t]$ as the velocity of GW $u$ at time $t$, which can be approximated by the discrete-time derivative $\dot{\mathbf{q}}_u[t] \approx \frac{\mathbf{q}_u[t+1] - \mathbf{q}_u[t]}{\Delta t}$. Consequently, each UAV is limited to a maximum speed $S_{\max}$ throughout the mission, restricting the instantaneous speed expressed as
\begin{equation}
\| \dot{\mathbf{q}}_u[t] \| \leq S_{\max}, \quad \forall t \in \mathcal{T} \setminus \{T\}.
\label{eq:velocity_constraint}
\end{equation}

Furthermore, UAVs must operate within the target area, ensuring that the position of each UAV remains within the limits given as
\begin{equation}
0 \le x_u[t] \le x^{\text{max}}, \quad \forall u \in \mathcal{U}, \; \forall t \in \mathcal{T},
\label{eq:x_max_constraint}
\end{equation}
\begin{equation}
0 \le y_u[t] \le y^{\text{max}}, \quad \forall u \in \mathcal{U}, \; \forall t \in \mathcal{T}.
\label{eq:y_max_constraint}
\end{equation}

Moreover, a minimum separation distance \( D_{\text{safe}} \) is enforced between two UAVs to prevent collisions. Therefore, for any pair of UAVs \( u \) and \( u' \), we have
\begin{equation}
    d_{u,u'}[t] \geq D_{\text{safe}}, \quad \forall u, u' \in \mathcal{U}, \; u \neq u', \; \forall t \in \mathcal{T},
\label{eq:safe_distance_constraint_uav}
\end{equation}
where \( d_{u,u'}[t] \) represents the Euclidean distance between the \( u \)-th UAV and the \( u' \)-th UAV at time \( t \) is expressed as $ d_{u,u'}[t] = \sqrt{(x_u[t] - x_{u'}[t])^2 + (y_u[t] - y_{u'}[t])^2}.$

\vspace{-0.5cm}
\subsection{Ground-to-Air (G2A) Communication Model}

In UAV communication systems, G2A link propagation plays a crucial role and has a significant impact on the communication quality between flying GWs and ground EDs \cite{zhou2022joint}. Given the dynamic nature of the environment, we employ a probabilistic path loss model to characterize G2A communication links. In this model, LoS and NLoS links are treated separately, with each link type associated with a distinct probability of occurrence. Accordingly, the probability of establishing a LoS connection between a GW \( u \) and ED \( v \) is given by
\begin{equation}
 P^{LoS}_{u,v}({\phi[t]})=\frac{1}{1+\vartheta e^{-\lambda(\phi_{u,v}[t]-\vartheta)}},
\label{eqt_p_los}
\end{equation}
where the parameters \(\vartheta\) and \(\lambda\) are constants determined by the environment, such as urban, dense urban, or rural settings. \(\phi[t]_{u,v}\) is the elevation angle from the \( u \)-th GW to the \( v \)-th ED at time \( t \), and it is expressed as
\begin{equation}
\phi_{u,v}[t] = \arctan\left(\frac{H_u}{\sqrt{(x_u[t] - x_v)^2 + (y_u[t] - y_v)^2}}\right).
\label{eq:elevation_angle}
\end{equation}

Furthermore, the probability of an NLoS is given by $P^{NLoS}_{u,v}({\phi[t]}) = 1 - P^{LoS}_{u,v}({\phi[t]})$. Consequently,  the total G2A path loss (in dB) between GW $u$ and ED $v$ at time $t$ is expressed as \cite{9310195}
\begin{equation}
\begin{split}
      L^{\text{G2A}}_{u, v}[t] = 20 \log_{10} \left( \frac{4\pi f d_{u,v}[t]}{c} \right) + \eta_{\text{LoS}}P^{LoS}_{u,v}(\phi[t])  \\ + \eta_{\text{NLoS}} P^{NLoS}_{u,v}(\phi[t]).
     \label{eqt_G2A_los}
\end{split}
\end{equation}

Here, $f$ is the carrier frequency in Hz, and $c$ is the speed of light in m/s. $\eta_{\text{LoS}}$ and $\eta_{\text{NLoS}}$ are environment-dependent constants that account for the additional average attenuation experienced under LoS and NLoS conditions, respectively.

\vspace{-3.5mm}
\subsection{UAV-ED Association Scheme}
In this work, each UAV is capable of serving multiple EDs simultaneously, enabling flexible and efficient data collection across the network. To manage this association, we define a binary indicator \( a_{u,v}[t] \in \{0, 1\} \), where \( a_{u,v}[t] = 1 \) indicates that the \( v \)-th ED is connected to the \( u \)-th UAV at time slot \( t \), and \( a_{u,v}[t]= 0 \) otherwise. Hence, we establish the following constraints
\begin{equation}
   a_{u,v}[t] = 
   \begin{cases} 
   1 & \text{if ED } v \text{ is connected to UAV } u, \\
   0 & \text{otherwise}.
   \end{cases}
   \label{eq:binary_constraint}
\end{equation}

To prevent an ED from connecting to multiple UAVs simultaneously, we impose a single-association constraint, expressed as
\begin{equation}
   \sum_{u \in \mathcal{U}} a_{u,v}[t] \le 1, \quad \forall v \in \mathcal{V}, \; \forall t \in \mathcal{T}.
   \label{eq:single_uav_per_ed}
\end{equation}

Next, we enforce a capacity constraint, where each UAV has a maximum number of EDs $\Lambda_{max}$ that a UAV can handle simultaneously, which is given by
\begin{equation}
   \sum_{v \in \mathcal{V}} a_{u,v}[t] \leq \Lambda_{max}, \quad \forall u \in \mathcal{U}, \; \forall t \in \mathcal{T},
   \label{eq:max_eds_per_uav}
\end{equation}

In a LoRa network, spreading factors are critical parameters that determine both data rate and transmission range. Therefore, in this work, the spreading factor for each uplink transmission is selected from a predefined set ${\bf \Psi} = \{7, 8, 9, 10, 11, 12\}$. For each ED $v$, a binary association variable $\psi_{v,n}[t]$ is define, where $n \in {\bf \Psi}$. Specifically, $\psi_{v,n}[t] = 1$ if ED $v$ communicates using spreading factor $n$ at time $t$, otherwise, $\psi_{v,n}[t] = 0$. Therefore, the selected spreading factor denoted by $\Psi_v[t] = \sum_{n \in {\bf \Psi}} \psi_{v,n}[t] \cdot n.$ Furthermore, we assume that each ED uses only one spreading factor at any given time $t$, and is given by
\begin{equation}
    \sum_{n \in \Psi} \psi_{v,n}[t] = 1, \quad \forall v \in \mathcal{V},\ t \in \mathcal{T}.
    \label{sf_binary}
\end{equation}

Furthermore, the transmission power level is selected from a predefined set $\mathcal{P} = \{p_{1}, \dots, p_{J}\}$, where each $p_j$ denotes a power level in dBm. For each ED $v$, we define a binary selection variable $p_{v,j}[t]$, where $j \in \mathcal{J} = \{1, \dots, J\}$. Specifically, $p_{v,j}[t] = 1$ if ED $v$ transmits with power level $p_j$ at time $t$, otherwise, $p_{v,j}[t] = 0$. The actual transmission power used by ED $v$ at time $t$ is then given by $
P_v[t] = \sum_{j \in \mathcal{J}} p_{v,j}[t] \cdot p_j$. To ensure that each ED uses only one transmission power level per time $t$, we impose the following constraint
\begin{equation}
    \sum_{j \in \mathcal{J}} p_{v,j}[t] = 1, \quad \forall v \in \mathcal{V},\ t \in \mathcal{T}.
    \label{tp_binary}
\end{equation}

Consequently, we define finite sets for all possible selections of spreading factors $\bar{\bf \Psi}$, transmission power allocations $\bar{\mathbf{P}}$, and  user associations $\mathbf{a}$, which can be expressed as
\begin{subequations}
\begin{align}
\bar{\mathbf{\Psi}} = \{ \Psi_{v}[t] \in {\bf \Psi} \mid \sum_{n=1}^{\Psi} \psi_{v,n}[t] \leq 1 , \quad \forall v \in \mathcal{V}\}. \label{sf} \\
 \bar{\mathbf{P}} = \{ P_{v}[t] \in \mathcal{P} |\sum_{j=1}^{\mathcal{J}} p_{v,j}[t] \leq 1 , \quad \forall v \in \mathcal{V}\}.  \label{tp} \\
\mathbf{a} = \{ a_{v}[t] \in \mathcal{U} |\sum_{u=1}^{\mathcal{U}} a_{u,v}[t] \leq 1 , \quad \forall v \in \mathcal{V}\}. \label{nodes}
\end{align}
\end{subequations}

\subsection{Duty-Cycle and Time-on-Air Constraint}
In the EU868 band, each ED is allowed to occupy the channel for only a small fraction $\eta_{\text{dc}}$ of the total time. Over a mission of duration $T_{\text{total}} = |\mathcal{T}|\,\Delta t$, this gives every ED an airtime budget $T_{\text{dc}} = \eta_{\text{dc}}\,T_{\text{total}}$. This budget matters because a LoRa packet does not have a fixed length, raising the SF $n$ doubles the symbol duration $T_{\text{sym}}(n) = 2^{n}/W$ and therefore lengthens the whole packet. A device on a high SF thus uses up its budget faster and can transmit less often. The time a single packet occupies the channel, its time-on-air, follows the LoRa specification~\cite{magrin2017performance} as $T_{\text{air}}(n) = \big(n_{\text{pre}} + 4.25 + N_{\text{pl}}(n)\big)\,T_{\text{sym}}(n)$, where $n_{\text{pre}}$ is the preamble length and $N_{\text{pl}}(n)$ the number of payload symbols given by $N_{\text{pl}}(n) = 8 + \max\!\big(\lceil \Phi(n)\rceil\,(\mathrm{CR}+4),\,0\big)$, with the header-and-coding term $\Phi(n) = \tfrac{8L - 4n + 28 + 16\,\mathrm{CRC} - 20H}{4\,(n - 2\,\mathrm{DE})}$. Here, $L$ is the payload size, $\mathrm{CR}\in\{1,2,3,4\}$ is the coding rate identifier, corresponding to a fractional coding rate of $4/(4{+}\mathrm{CR})$, and $H$ is the header indicator. Therefore, the budget is enforced on the airtime each ED accumulates over the mission, which is expressed as
\begin{equation}
\sum_{t \in \mathcal{T}} a_{u,v}[t]\, T_{\text{air}}\!\big(\Psi_v[t]\big) \le T_{\text{dc}},
\qquad \forall v \in \mathcal{V}.
\label{eq:duty_cycle}
\end{equation}

\subsection{Energy Efficiency Model}\label{ee_model}
To link our network topology with the LoRa physical layer parameters, we begin by modeling the signal-to-noise ratio (SNR) between UAV \( u \) and ED \( v \) at time \( t \), which is given by
\begin{equation}
   \rho_{u,v}[t] = \frac{P_{v}[t] \cdot G^{\text{eff}}_{u,v}[t]}{\sigma^{2}},
   \label{eq:snr}
\end{equation}
where \( P_{v}[t] \) is the transmission power of ED \( v \). To reflect practical conditions, the effective channel gain combines the large-scale gain with small-scale fading, that is, $G^{\text{eff}}_{u,v}[t] = G_{u,v}[t]\,|h_{u,v}[t]|^2$, with $G_{u,v}[t] = 10^{-\frac{1}{10} L^{\text{G2A}}_{u,v}[t]}$ and $|h_{u,v}[t]|^2$ a unit-mean Rician fading power drawn per link per slot. Its $K$-factor is elevation-dependent, such that $\kappa_{u,v}[t] = A_1 e^{A_2 \phi_{u,v}[t]}$. The noise power $\sigma^{2}$ in Eq.\eqref{eq:snr} is obtained from the receiver noise floor $-174 + 10\log_{10}(W) + NF$ [dBm], where $NF$ is the receiver noise figure.

All flying GWs share a LoRa band of bandwidth $W$. Because SFs are only quasi-orthogonal, a transmission suffers interference from both co-SF and inter-SF~\cite{waret2018lora}. In this work, we enforce that an uplink packet from ED $v$ to GW $u$ at SF $n$ is received only if it satisfies three conditions. First, the received SNR must exceed the per-SF reception threshold given as $\rho_{u,v}[t] \ge \bar{\Omega}_{n}$. Second and third, it must win the capture contest against each interferer class $\varpi \in \{\mathrm{co},\mathrm{in}\}$, which is expressed as
\begin{equation}
\Omega^{\varpi}_{u,v}[t] = \frac{\rho_{u,v}[t]}{\sum_{v' \in \mathcal{I}^{\varpi}_{u,v}[t]} \rho_{u,v'}[t] + 1} \ \ge\ q^{\varpi},
\quad \varpi \in \{\mathrm{co},\mathrm{in}\},
\label{eq:capture}
\end{equation}
where $\mathcal{I}^{\mathrm{co}}_{u,v}[t]$ and $\mathcal{I}^{\mathrm{in}}_{u,v}[t]$ collect the in-range EDs that share and differ from $v$'s SF, respectively. The thresholds $\bar{\Omega}_{n}$ and the per-SF $q^{\mathrm{in}}_{n}$ are given in Table~\ref{tab:thresholds} while $q^{\mathrm{co}}=6$~dB.

Additionally, LoRa access is pure ALOHA, so overlapping packets may still collide. A tagged packet facing $N$ in-range interferers of one class is lost with probability $P_{ovl}(N)=1-e^{-2N\eta_{dc}}$, unless it has already cleared that class's capture. The two classes combine into the delivery probability, given as
\begin{equation}
P^{deliv}_{u,v}[t] = \big(1 - P^{co}_{loss}\big)\big(1 - P^{in}_{loss}\big),
\label{eq:deliver}
\end{equation}
where $P^{co}{loss}$ is zero if the co-SF capture condition holds and equals $P{ovl}(N^{co}{u,v}[t])$ otherwise, while $P^{in}{loss}$ is zero if the inter-SF capture condition holds and equals $P_{ovl}(N^{in}_{u,v}[t])$ otherwise. Given the probability of delivery $P^{deliv}_{u,v}[t]$, the achievable data rate for the link between ED $v$ and GW $u$ at time $t$ is calculated as
\begin{equation}
\Re_{u,v}[t] = P^{deliv}_{u,v}[t]\, W \log_2\!\big(1 + \rho_{u,v}[t]\big).
\label{eq:rate}
\end{equation}

Furthermore, we quantify the uplink communication EE of UAV $u$ as the aggregate uplink data rate of its associated EDs per unit of power consumed by these devices. Accordingly, the uplink EE of UAV $u$ at time slot $t$, is expressed as
\begin{equation}
\zeta_{u}[t] = \frac {\sum_{v \in \mathcal{V}} a_{u,v}[t] \Re_{u,v}[t]}{P_{T} + P_{c}},
\label{eqt_EE}
\end{equation}
where \( P_{c} = \sum_{v \in \mathcal{V}} a_{u,v}[t] P_{\text{circuit},v} \) is the constant circuit power of associated EDs, and \( P_{T} = \sum_{v \in \mathcal{V}} a_{u,v}[t] P_{v}[t] \) is the total uplink transmission power. The UAV propulsion power (detailed in the next subsection) is deliberately excluded from the denominator of Eq.~\eqref{eqt_EE} for two reasons. First, the objective targets uplink communication efficiency, which matters most for energy-constrained LoRa EDs. Second, propulsion power would dominate the denominator and render $\zeta_{u}[t]$ insensitive to EE optimization. Instead, it is handled as a physical constraint bounded by $P_{\text{max}}$, as formulated in the following section. Since our main objective is to maximize the system's weighted energy efficiency $EE_{sys}$, we express it as
\begin{equation}
    EE_{sys} = \sum_{t=1}^{T} \sum_{u=1}^{U} \lambda_u \cdot \zeta_{u}[t],
    \label{weighted_ee}
\end{equation}
where $\lambda_u \ge 0$ is the weight assigned to UAV $u$, with $\sum_{u} \lambda_u = 1$. These weights can prioritize gateway tasks, for instance to balance energy across the fleet or to cover critical areas. In this work, the UAVs are identical and symmetrically deployed, so we set $\lambda_u = 1/U$, and $EE_{sys}$ then reflects the average per-gateway efficiency without favoring any particular UAV.

\subsection{Power Consumption Model for a Single-Rotor UAV}
In this work, we adopt a power consumption model for a single-rotor UAV following the approach in \cite{gong2023modeling}. Specifically, we first consider the total hover power $P_{1}$, which is expressed as
\begin{equation}
    P_{1}
    \;=\;
    \underbrace{\frac{\bar{\delta} \,\rho \, s \, A \,\Omega^{3} \, R^{3}}{8}}_{P_{0}}
    \;+\;
    \underbrace{(1 + k)\,\frac{W^{3/2}}{\sqrt{2\,\rho\,A}}}_{P_{i}},
    \label{eq:p1}
\end{equation}
where $P_{0}$ represents the blade profile power, and $P_{i}$ is the induced power. To account for forward flight, the power consumption $P_{\text{prop}}(\dot{\mathbf{q}}_u[t])$ is modeled as follows
\begin{equation}
    \begin{split}
            P_{\text{prop}}(\dot{\mathbf{q}}_u[t]) = \; P_{0}  \left( 1 + \frac{3 \dot{\mathbf{q}}_u[t]^{2}}{\Omega^{2} R^{2}} \right) +   \frac{1}{2}\,S_{F\parallel} \rho \dot{\mathbf{q}}_u[t]^{3} + \\  
            P_{i}\,\tilde{\kappa}\, \left( \sqrt{\tilde{\kappa}^{2} +  \tfrac{\dot{\mathbf{q}}_u[t]^{4}} {4\,v_{0}^{4}} } - \tfrac{\dot{\mathbf{q}}_u[t]^{2}}{2\,v_{0}^{2}} \right),
    \label{eq:prop_uav}
    \end{split}
\end{equation}
where $\dot{\mathbf{q}}_u[t]$ denotes the forward speed at time $t$, $\bar{\delta}$ is the profile drag coefficient, $\rho$ is the air density (kg/m$^3$), $s$ is the rotor solidity, $A$ is the rotor disc area ($m^2$), $\Omega$ is the blade angular velocity (rad/s), $R$ is the rotor radius ($m$), $W$ is the UAV weight (in newtons), $k$ is the induced power correction factor, $\tilde{\kappa}$ is the thrust-to-weight ratio. Unless otherwise stated, the parameters used in this work are based on \cite{zeng2019energy}.

Furthermore, we impose a maximum power budget \( P_{\text{max}} \) that the UAV must not exceed at any given time \( t \) and is expressed as
\begin{equation}
    P_{\text{prop}}(\dot{\mathbf{q}}_u[t]) \leq P_{\text{max}}, \quad \forall u \in  \mathcal{U}, \forall t \in  \mathcal{T}.
    \label{eq:prop_uav_max}
\end{equation}

\vspace{-0.3cm}
\section{Problem Formulation} \label{Section_4} 
In the following, we aim to maximize the system's weighted EE of flying LoRa GWs as defined in Eq.~\eqref{eqt_EE}. To achieve this objective, we jointly optimize the followings, (i) trajectories of the UAVs, denoted by \( \mathbf{Q} = \{ \mathbf{q}_u[t] \}_{u \in \mathcal{U},\ t \in \mathcal{T}} \), where each trajectory point \( \mathbf{q}_u[t] \) is determined by the UAV's movement speed \( s_u[t] \) and movement direction \( \theta_u[t] \), (ii) the LoRa communication parameters, specifically the finite sets of spreading factors $\bar{\mathbf{\Psi}}$ and transmission powers $\bar{\mathbf{P}}$, which directly affect the achievable data rates of the EDs, and (iii) the ED associations to balance traffic load and significantly improve overall network performance. Consequently, our optimization problem is mathematically formulated as follows
\begin{subequations}\label{formulated_problem_1}
\begin{align}
\left( \textbf{P1} \right) \quad \;\;
& \max_{\mathbf{Q}, \bar{\mathbf{\Psi}},  \bar{\mathbf{P}}, a }  \qquad EE_{sys}, \label{constr:a} \\
\text{s.t.} \quad \;\;
& P_v[t] \in \mathcal{P}, \forall v \in \mathcal{V}, \forall t \in \mathcal{T}, \label{constr:b} \\
& \;a_{u,v}[t] \in \{0, 1\}, \forall u \in \mathcal{U}, \forall t \in \mathcal{T}, \label{constr:c} \\
& \;\sum_{u \in \mathcal{U}} a_{u,v}[t] \leq 1, \forall v \in \mathcal{V}, \forall t \in \mathcal{T}, \label{constr:d} \\ 
& \;\psi_{v}[t] \in {\bf \Psi}, \forall v \in \mathcal{V}, \forall t \in \mathcal{T}, \label{constr:f} \\
& \; \rho_{u,v}[t] \ge \bar{\Omega}_{\Psi_v[t]}, \forall v \in \mathcal{V}, \forall t \in \mathcal{T}, \label{constr:g} \\
& \; \eqref{eq:position_update_uav}, \eqref{eq:velocity_constraint}, \eqref{eq:x_max_constraint}, \eqref{eq:y_max_constraint}, \eqref{eq:safe_distance_constraint_uav}, \eqref{eq:max_eds_per_uav}, \eqref{eq:duty_cycle}, \text{ and } \eqref{eq:prop_uav_max}.
\end{align}
\end{subequations}
\begin{table}[t]
    \centering
    \caption{\small Per-SF Reception and Inter-SF Capture Thresholds with $W = 125$ kHz~\cite{waret2018lora}}
    \label{tab:thresholds}
    {\footnotesize
    \begin{tabular}{|c|c|c|c|c|c|c|}
        \hline
        SF & 7 & 8 & 9 & 10 & 11 & 12 \\
        \hline
        $\bar{\Omega}_{n}$ (dB) & $-6$ & $-9$ & $-12$ & $-15$ & $-17.5$ & $-20$ \\
        \hline
        $q^{in}_{n}$ (dB) & $-7.5$ & $-9$ & $-13.5$ & $-15$ & $-18$ & $-22.5$ \\
        \hline
    \end{tabular}
    }
\end{table}
Here, constraint~\eqref{constr:b} restricts the transmission power to a predefined discrete set. The binary association indicator $a_{u,v}[t]$ in constraint~\eqref{constr:c} ensures that each ED is associated with at most one GW. Constraint~\eqref{constr:d} further ensures that each ED accesses at most one channel at any time. Furthermore, constraint~\eqref{constr:f} defines the selection of spreading factors from the available set. Finally, constraint~\eqref{constr:g} enforces the per-SF reception condition, requiring the received SNR to exceed the threshold $\bar{\Omega}_{n}$, where $n = \Psi_v[t]$ is the SF selected by ED $v$, listed in Table~\ref{tab:thresholds} given by~\cite{waret2018lora}, while successful reception additionally requires the co-/inter-SF capture conditions of Eq.~\eqref{eq:capture}.

Given the multiple objectives of our work defined earlier, we need to reformulate the original problem (P1) in Eq.~\eqref{formulated_problem_1} to handle several key aspects, such as encouraging agents to associate with as many EDs as possible, avoiding collisions with other flying GWs, and arriving at the final destination. Therefore, P1 can be reformulated as
\begin{subequations}\label{formulated_problem_2}
\begin{align}
\left( \textbf{P2} \right)
& \max_{\mathbf{Q}, \bar{\boldsymbol{\Psi}}, \bar{\mathbf{P}}, a } \;
\sum_{t \in \mathcal{T}} \Big(
\omega_1 c_u[t] + \omega_2 \nu_u[t] + \omega_3 EE_{\text{sys}}[t] \nonumber \\
& \qquad\qquad + \omega_4 \xi_u[t] - \omega_5 C_u[t] - \omega_6 d_{cs,u}[t] \Big), \label{constr:multi_obj} \\
\text{s.t.} \quad
& \eqref{constr:b}, \eqref{constr:c}, \eqref{constr:d}, \eqref{constr:f}, \eqref{constr:g}, \nonumber \\
& \eqref{eq:position_update_uav}, \eqref{eq:velocity_constraint}, \eqref{eq:x_max_constraint}, \eqref{eq:y_max_constraint}, \eqref{eq:safe_distance_constraint_uav}, \eqref{eq:max_eds_per_uav}, \eqref{eq:duty_cycle}, \text{ and } \eqref{eq:prop_uav_max}.
\end{align}
\end{subequations}

Here, $\omega_1, \dots, \omega_6$ are tunable weights, $c_u[t]$ denotes the decode success incentive, $\nu_u[t]$ is the first-time coverage term, $\xi_u[t]$ represent the inter-UAV spread incentive, the collision penalty is denoted as $C_u[t]$, and the distance to the CS $d_{cs,u}[t] = \| \mathbf{q}_u[t] - \mathbf{q}_{cs} \|$. 

It should be noted that the optimization problem in \eqref{formulated_problem_2} is known to be NP-hard and non-convex, owing to the coupling of binary ED-association variables with high-dimensional, time-varying communication and trajectory parameters~\cite{su2020energy}. Conventional approaches, such as convex relaxation and alternating optimization, generally rely on approximations or problem decomposition and require the resulting optimization problem to be repeatedly solved as the network conditions change. This may introduce considerable computational complexity and lead to suboptimal solutions, particularly in the decentralized, partially observable, and dynamic setting considered in this work. DRL is well suited to this setting because it can learn a sequential decision-making policy that captures the interactions among the jointly optimized variables under dynamic and partially observable network conditions. Once trained, the learned policy enables low-latency decisions at each time step without solving the original optimization problem from scratch whenever the network conditions change. This motivates the learning-based formulation presented in the following section.

\section{MARL-based Resource Allocation} \label{Section_5}
To solve P2, we transform the optimization problem into a multi-agent sequential decision-making problem in which each UAV, acting under partial observability, uses its local observations to drive actions that improve the global network state and system energy efficiency. Unlike convex relaxation or alternating optimization, MARL learns joint policies over the coupled action space directly through environmental interaction. Consequently, we adopt the CTDE paradigm, which lets agents exploit global information during training while executing on purely local observations, thereby handling the coordination and partial-observability challenges that neither centralized solvers nor independent single-agent methods can address simultaneously. Thus, we model this interaction as a POSG~\cite{albrecht2024multi} and solve it with MAPPO~\cite{yu2022surprising}. Formally, the POSG is defined by the tuple $\langle \mathcal{U}, \mathcal{S}, \{\mathcal{A}_u\}, \{\mathcal{O}_u\}, \mathcal{T}, \{\mathcal{Z}_u\}, \{\mathcal{R}_u\}, \gamma, \mu_0 \rangle$, whose elements are described as follows.

\subsubsection{Global State Space $\{\mathcal{S}\}$}
The global state $\mathbf{s}[t] \in \mathcal{S}$ aggregates the full network information available to the centralized critic during training. It comprises the positions $\{\mathbf{q}_u[t]\}_{\forall u \in \mathcal{U}}$, velocities $\{\dot{\mathbf{q}}_u[t]\}_{\forall u \in \mathcal{U}}$, and propulsion-power states $\{\kappa_u[t]\}_{\forall u \in \mathcal{U}}$ of all UAVs, as well as the positions $\{\mathbf{g}_v\}_{\forall v \in \mathcal{V}}$ of all EDs, the full channel-gain matrix $\{G^{\text{eff}}_{u,v}[t]\}_{\forall u \in \mathcal{U}, \forall v \in \mathcal{V}}$, and the charging-station location $\mathbf{q}_{cs}$. Thus, the global state is denoted as $\mathbf{s}[t] = \big\{ \{\mathbf{q}_u[t]\},\ \{\dot{\mathbf{q}}_u[t]\},\ \{\kappa_u[t]\},\ \{\mathbf{g}_v\},\ \{G^{\text{eff}}_{u,v}[t]\},\ \mathbf{q}_{cs} \big\}$. 

\subsubsection{Action Space $\{\mathcal{A}_u\}$}
At time $t$, agent $u$ selects $a_u[t]\in\mathcal{A}_u$, comprising mobility and resource-allocation variables. Therefore, the individual action of agent $u$ at time $t$ is given by
\begin{equation}
    a_u[t] = \big\{\theta_u[t],\ s_u[t],\ \boldsymbol{\psi}_u[t],\ \mathbf{p}_u[t]\big\},
    \label{eq:mixed_action}
\end{equation}
where $\theta_u[t]$ and $s_u[t]$ denote the movement direction and speed of UAV $u$, respectively, while $\boldsymbol{\psi}_u[t] = \{\psi_{u,k}[t]\}_{k=1}^{\Lambda_{\max}}$ and $\mathbf{p}_u[t] = \{p_{u,k}[t]\}_{k=1}^{\Lambda_{\max}}$ contain the SF and TP selections. Since mobility is inherently continuous, whereas SF and TP are selected from finite sets, we adopt a mixed action space. Specifically, $\theta_u[t]\in[-\beta/2,\beta/2]$ and $s_u[t]\in[0,S_{\max}]$ are continuous, while $\boldsymbol{\psi}_u[t]$ and $\mathbf{p}_u[t]$ are discrete.  Furthermore, the joint action space is given by the Cartesian product $\mathcal{A} = \prod_{u \in \mathcal{U}} \mathcal{A}_u$, and the joint action at time $t$ is expressed as $\mathbf{a}[t] = (a_1[t], a_2[t], \ldots, a_U[t]) \in \mathcal{A}$.

\subsubsection{Observation Space $\{\mathcal{O}_u\}$}
Due to the dynamic nature of our environment, driven by high UAV mobility and a time-varying wireless medium, each UAV $u$ operates under partial observability with its own observation space $\mathcal{O}_u$. An agent's policy $\pi_u$ uses this observation $o_u[t]$ as input to produce an action $a_u[t]$. Specifically, UAV $u$ at time $t$ has access only to its local state and sensed environmental information. The observation $o_u[t] \in \mathcal{O}_u$ of UAV $u$ is defined as
\begin{equation}
\begin{aligned}
o_u[t] = \big\{\, &\mathbf{q}_u[t],\ \dot{\mathbf{q}}_u[t],\ \kappa_u[t],\ \big(\theta_u[t-1], s_u[t-1]\big),\ \Delta\mathbf{q}^{cs}_u[t], \\
&\{\, \mathbf{g}_v[t],\ d_{u,v}[t],\ G^{\text{eff}}_{u,v}[t] \mid \forall v \in \mathcal{V}_u[t] \,\}, \\
&\{\, \mathbf{q}_{u'}[t] - \mathbf{q}_u[t],\ d_{u,u'}[t] \mid \forall u' \in \mathcal{N}_u[t] \,\} \,\big\}.
\end{aligned}
\label{eq_agent_observation}
\end{equation}

Here, $\dot{\mathbf{q}}_u[t]$ is the UAV's velocity, $\kappa_u[t]$ its instantaneous propulsion power consumption, $\theta_u[t-1]$ and $s_u[t-1]$ are the continuous movement direction and speed applied at the previous time step. $\Delta\mathbf{q}^{cs}_u[t] \triangleq \mathbf{q}_{cs} - \mathbf{q}_u[t]$ is the displacement vector from the UAV to the charging station. Furthermore, the neighbor set $\mathcal{N}_u[t] = \{\, u' \neq u : \|\mathbf{q}_{u'}[t] - \mathbf{q}_u[t]\| \le R_{\text{sense}} \,\}$ contains all other UAVs within the on-board radar sensing range $R_{\text{sense}} \gg D_{\text{safe}}$, for which UAV $u$ observes their relative positions $\mathbf{q}_{u'}[t] - \mathbf{q}_u[t]$ and separations $d_{u,u'}[t]$. This component is essential for the agent to learn the collision-avoidance behavior by sensing neighbors well before the safety threshold $D_{\text{safe}}$ is reached, and the agent retains sufficient horizon to adjust its trajectory. Therefore, the joint observation\footnote{To guarantee a fixed-size input to the neural network across varying numbers of in-range EDs and neighboring UAVs, two zero-padding mechanisms are used. First, virtual EDs are appended when fewer than the maximum expected EDs are within communication range. Second, the neighbor-UAV slots are zero-padded when fewer than $U-1$ UAVs lie within the sensing range $R_{\text{sense}}$.} at time $t$ is denoted as $\mathbf{o}[t] = (o_1[t], o_2[t], \ldots, o_U[t])$.

\subsubsection{State Transition Function $\mathcal{T}$}
The state transition function $\mathcal{T}: \mathcal{S} \times \mathcal{A} \times \mathcal{S} \rightarrow [0,1]$ gives the probability $\mathcal{T}(\mathbf{s}[t], \mathbf{a}[t], \mathbf{s}[t+1]) = P(\mathbf{s}[t+1] \mid \mathbf{s}[t], \mathbf{a}[t])$ of transitioning to $\mathbf{s}[t+1]$ under the joint action $\mathbf{a}[t]$, following the Markovian property. Its stochasticity arises mainly from the time-varying channel, as the per-slot Rician fading and ALOHA-based reception randomize the channel-gain component of the next state, while the UAV mobility evolves deterministically under the joint action.

\subsubsection{Observation Function $\{\mathcal{Z}_u\}$}
For each agent $u$, the observation function $\mathcal{Z}_u: \mathcal{S} \times \mathcal{O}_u \rightarrow [0,1]$ specifies the probability $\mathcal{Z}_u(\mathbf{s}[t], o_u[t]) = P(o_u[t] \mid \mathbf{s}[t])$ of receiving observation $o_u[t]$ given the global state $\mathbf{s}[t]$. This captures the partial-observability constraint, where $o_u[t]$ is a localized subset of $\mathbf{s}[t]$. Therefore, the agent $u$ perceives only the EDs in-range within its communication range $R_{comm}$, rather than all of them.

\subsubsection{Reward Function $\{\mathcal{R}_u\}$}
At each time $t$, the instantaneous reward for agent $u$ is given by $\mathcal{R}_u: \mathcal{S} \times \mathcal{A} \rightarrow \mathbb{R}$, which may depend on the global state and the joint action of all agents. It is defined as
\begin{equation}
\begin{aligned}
   \mathcal{R}_u(\mathbf{s}[t], \mathbf{a}[t]) = & \; \omega_1 {c}_u[t] + \omega_{2}\, \nu_u[t] + \omega_3 EE_{\text{sys}}[t] + \omega_{4}\, \xi_u[t] \\
   & - \omega_5 C_u[t] - \omega_6 d_{cs,u}[t] - \mathbb{I}_{[t = T]}\, \omega_7 (1 - \Xi_u^{cs}),
\end{aligned}
\label{eq:step_reward}
\end{equation}
where ${c}_u[t] \in \{0, 1\}$ equals 1 if UAV $u$ successfully decodes at least one uplink transmission from a served ED at time $t$, and 0 otherwise. The term $\nu_u[t]$ counts the EDs served by UAV $u$ for the first time in the episode, providing a coverage incentive that steers UAVs toward as-yet-unserved devices. $EE_{\text{sys}}[t]$ is the instantaneous global system energy efficiency. The term $\xi_u[t] = \min\!\big(\delta_u[t]/\delta_{\text{cap}},\, 1\big)\in[0,1]$ is a saturating inter-UAV spread incentive, where $\delta_u[t]$ is the distance from UAV $u$ to its nearest neighbor and $\delta_{\text{cap}}$ a saturation distance. Note that this encourages the fleet to fan out and broaden spatial coverage. The indicator $C_u[t] \in \{0, 1\}$ marks a collision, i.e., a separation below $D_{\text{safe}}$ from another UAV, and $d_{cs,u}[t] = \| \mathbf{q}_u[t] - \mathbf{q}_{cs} \|$ is the distance from UAV $u$ to the CS. The final term applies a terminal penalty via $\mathbb{I}_{[t=T]}$, where $\Xi_u^{cs} \in \{0,1\}$ equals 1 if UAV $u$ reaches the CS, so $\omega_7(1 - \Xi_u^{cs})$ is incurred only on failure to arrive. An episode terminates for a UAV upon an inter-UAV collision, a boundary violation, or arrival at the CS. Note that the first two are unsafe terminations whose penalties discourage infeasible trajectories, while the third marks successful mission completion. 
\begin{algorithm}[t!]
\caption{Gain-Based UAV-ED Association}
\label{alg:gain_based_association}
\begin{algorithmic}[1]
\State \textbf{Input:} UAVs $\mathcal{U}$, EDs $\mathcal{V}$, positions $\mathbf{q}_u[t]$ and $\mathbf{g}_v$, max quota $\Lambda_{\max}$, communication range $R_{\text{comm}}$, gain matrix $G_{u,v}[t]$
\State \textbf{Output:} Association set $\mathcal{\hat{A}}$
\State Initialize $\mathcal{\hat{A}} \gets \emptyset$, capacity counter ${c}_u \gets 0$ for all $u$
\For{each $v \in \mathcal{V}$}
    \State Set $\mathcal{U}_v \gets \{ u \in \mathcal{U} \mid \| \mathbf{q}_u[t] - \mathbf{g}_v \| \le R_{\text{comm}} $
    \State and ${c}_u < \Lambda_{\max} \}$
    \If{$\mathcal{U}_v \ne \emptyset$}
        \State $u^* = \arg\max_{u \in \mathcal{U}_v} G_{u,v}[t]$
        \State Add $(u^*, v)$ to $\mathcal{\hat{A}}$, set $\hat{a}_{u^*} \gets \hat{a}_{u^*} + 1$
    \EndIf
\EndFor
\State \Return $\mathcal{\hat{A}}$
\end{algorithmic}
\end{algorithm}

The weights $\{\omega_1, \dots, \omega_7\}$ are tunable hyperparameters controlling the relative importance of each component. They are tuned so that no objective dominates the policy gradient. For instance, the energy-efficiency weight $\omega_3$ is kept small to offset the large numerical range of $EE_{\text{sys}}$, while the coverage and terminal-arrival weights are set larger to make first-time service and mission completion salient. This balanced design lets the agents jointly trade off coverage, energy efficiency, safety, and mission completion. Therefore, the total cooperative reward across all $U$ UAVs is $r[t] = \sum_{u=1}^{U} \mathcal{R}_u(\mathbf{s}[t], \mathbf{a}[t])$.

It should be noted that the discount factor $\gamma \in (0,1)$ balances immediate and long-term rewards, prioritizing near-future rewards over distant ones. The initial state distribution $\mu_0: \mathcal{S} \rightarrow [0,1]$ specifies the probability distribution over initial states, where $\mathbf{s}_0 \sim \mu_0(\cdot)$ determines the starting configuration of UAV positions, ED locations, and initial network parameters. Furthermore, each UAV agent $u$ follows a stochastic policy $\pi_u: \mathcal{O}_u \times \mathcal{A}_u \rightarrow [0,1]$, parameterized by $\phi_u$, which maps its local observation to a probability distribution over actions. The joint policy is denoted as $\boldsymbol{\pi} = (\pi_1, \pi_2, \ldots, \pi_U)$. Therefore, the objective of each agent is to maximize its expected cumulative discounted reward given as
\begin{equation}
J_u(\pi_u) = \mathbb{E}_{\tau_u} \left[ \sum_{t=0}^{T} \gamma^t \mathcal{R}_u(\mathbf{s}[t], \mathbf{a}[t]) \right],
\end{equation}
where the expectation is taken over trajectories $\tau_u$ of agent $u$. In the cooperative setting, agents aim to maximize the team objective $J(\boldsymbol{\pi}) = \sum_{u=1}^{U} J_u(\pi_u)$.

\section{GLo-MAPPO for Intelligent UAV-Assisted LoRa Networks} \label{Section_6}
We now present GLo-MAPPO, a gain and energy-aware MARL framework for UAV-assisted LoRa networks. It has two components, namely, a gain-based association step that assigns each ED to the most suitable UAV, and a MARL training loop that optimizes trajectories and communication parameters. We describe each in turn and then combine them to solve P2~\eqref{formulated_problem_2}.

\subsection{Gain-Based Association Algorithm}
For each ED, the algorithm links it to the in-range UAV with the strongest channel, provided that UAV still has quota capacity. The gain $G^{\text{eff}}_{u,v}[t] = G_{u,v}[t]\left|h_{u,v}[t]\right|^{2}$ reuses the path loss from Eq.~\eqref{eqt_G2A_los} and the Rician fading power $\left|h_{u,v}[t]\right|^{2}$. Therefore, the pairwise gains form the matrix $\mathbf{G}^{\text{eff}}[t] \in \mathbb{R}^{U \times V}$, which is given by
\begin{equation}
\mathbf{G}^{\text{eff}}[t] =
\begin{bmatrix}
G^{\text{eff}}_{1,1}[t] & \cdots & G^{\text{eff}}_{1,V}[t] \\
\vdots & \ddots & \vdots \\
G^{\text{eff}}_{U,1}[t] & \cdots & G^{\text{eff}}_{U,V}[t]
\end{bmatrix}.
\label{eq:gain_matrix}
\end{equation}

Additionally, the algorithm iterates over each ED to determine the set of eligible UAVs, and selects the one that offers the maximum channel gain. If no eligible UAV exists for a particular ED, it remains unassociated at that time step. Moreover, to ensure the system adapts to the high mobility of our flying UAVs, the association scheme in Algorithm~\ref{alg:gain_based_association} is updated at every timestep $t$. This update occurs on a timescale of $\Delta t$, meaning that as the UAV trajectories $\mathbf{Q}$ evolve, the association is recalculated to maintain the strongest possible links. The details of our association scheme are shown in Algorithm~\ref{alg:gain_based_association}.
\begin{algorithm}[t]
\caption{GLo-MAPPO Training Algorithm}
\label{alg:glo_mappo_algo}
\begin{algorithmic}[1]
\State \textbf{Input:} Number of agents $U$, learning rate $\alpha$, clipping range $\epsilon$, discount factor $\gamma$, target update rate $\tau$
\State \textbf{Initialize:} Shared policy network $\pi_\phi$, Value network $V_\mu$ with random parameters $\phi$, $\mu$.
\State \textbf{Output:} Optimized UAV trajectory, TP, and SF allocation
\For{each episode $e = 1, \dots, e_{\max}$}
    \State Initialize state $\mathbf{s}[0]$ and observations $\{o_u[0]\}_{u=1}^{U}$ for all
    \State agents
    \For{each time step $t = 0, \dots, T-1$}
        \For{each agent $u = 1, \dots, U$}
            \State Sample action $a_u[t]\sim\pi_{\phi}(\cdot\mid o_u[t],\mathbf{e}_u)$
            \State Perform association using \textbf{Algorithm}~\ref{alg:gain_based_association}
            \If{$a_u[t]$ violates constraints}
                \State Reinitialize $a_u[t]$ to satisfy constraints
            \Else
                 \State Execute $a_u[t]$, observe $o_u[t+1]$ and reward \State $\mathcal{R}_u(\mathbf{s}[t],\mathbf{a}[t])$
            \EndIf
        \EndFor
        \State Store transitions $\{\mathbf{s}[t], \mathbf{a}[t], r_u[t], \mathbf{s}[t+1]\}$ in $\mathcal{D}$
    \EndFor
    \For{epoch $k = 1, \dots, K$}
        \State Compute advantage $\{\hat{A}_u[t]\}_{u=1}^{U}$ using Eq.~\eqref{eq:advantage_function}
        \State Update $\phi$ with PPO clipped objective in Eq.~\eqref{eq:actor_loss}
        \State Update $\mu$ by minimizing the mean square error in 
        \State Eq.~\eqref{eq:critic_loss}
    \EndFor
    \State Periodically evaluate policy performance and adjust
    \State hyperparameters
\EndFor
\State \textbf{Return:} Optimized policies $\{\pi_{\phi_u}\}_{u=1}^{U}$ and $V_\mu$
\end{algorithmic}
\end{algorithm}
\vspace{-1.0em}

\begin{figure}[t]
  \centering
  \includegraphics[width=0.7\linewidth]{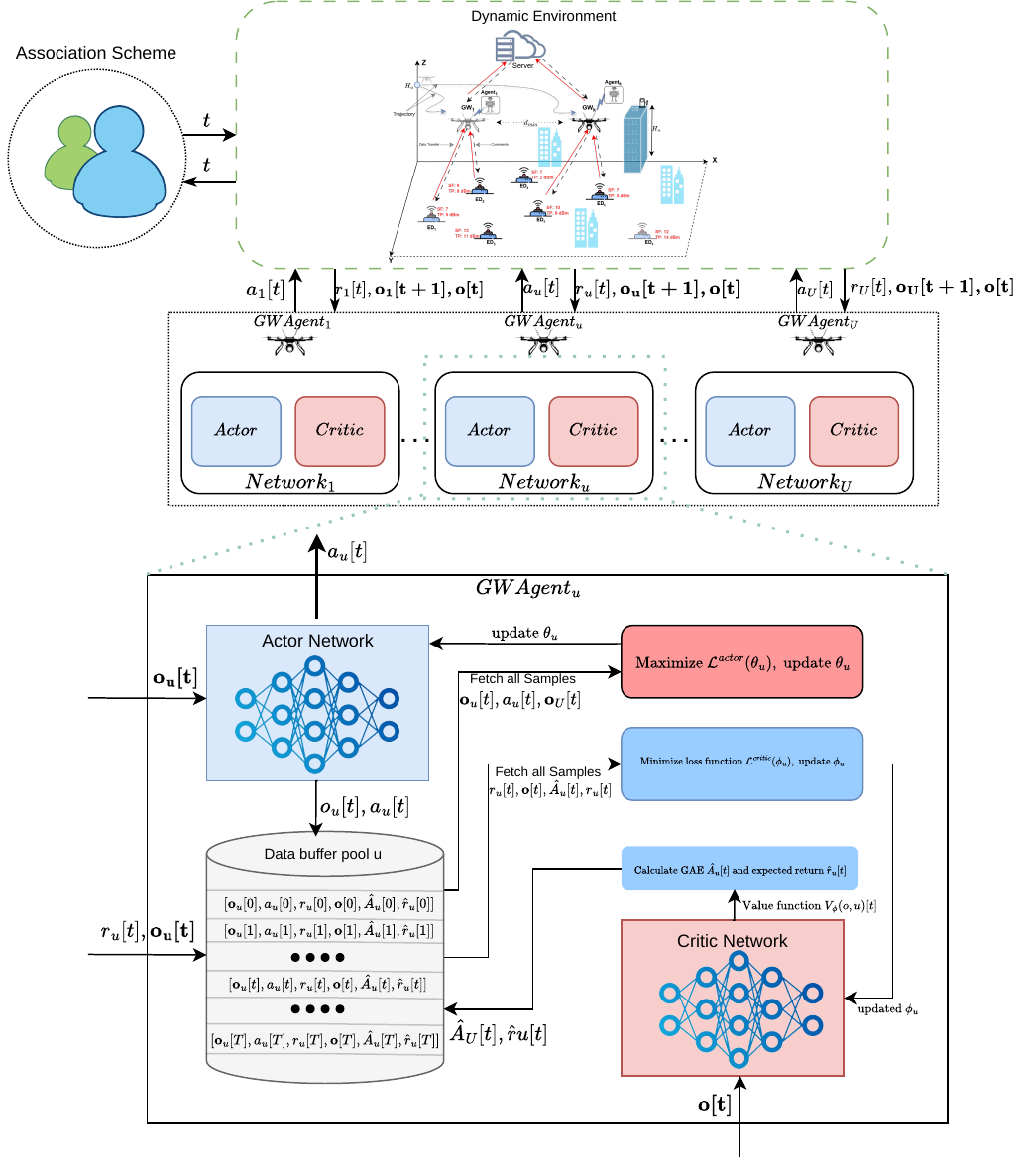}
  \caption{GLo-MAPPO system architecture.}
  \label{fig_GLo_MAPPO_Architecture}
  \vspace{-1.0em}
\end{figure}

\subsection{Proposed GLo-MAPPO Algorithm}\label{section_glo_mappo_algo_B}
As shown in Fig.~\ref{fig_GLo_MAPPO_Architecture}, our proposed GLo-MAPPO architecture follows the CTDE paradigm. Specifically, all agents share a policy network $\pi_\phi$, conditioned on a one-hot identifier $\mathbf{e}_u$, which maps each local observation $o_u[t]$ to the corresponding control variables. A shared centralized value network $V_\mu$ is trained using the global state $\mathbf{s}[t]$, whereas each agent relies solely on $o_u[t]$ during execution. he advantages are estimated from an $n$-step bootstrapped return using the target value network $V_{\mu^{\mathrm{old}}}$. For agent $u$, the advantage function $\hat{A}_u[t]$ can be expressed as
\begin{equation}
\hat{A}_u[t] = \hat{R}_u[t] - V_\mu(\mathbf{s}[t]),
\label{eq:advantage_function}
\end{equation}
where $\hat{R}_u[t] = \sum_{l=0}^{n-1}\gamma^l \mathcal{R}_u(\mathbf{s}[t+l],\mathbf{a}[t+l]) + \gamma^n V_{\mu^{\mathrm{old}}}(\mathbf{s}[t{+}n])$, and $n$ denotes the return horizon. To handle the mixed action space defined in Eq. \eqref{eq:mixed_action}, the actor processes $[o_u[t],\mathbf{e}_u]$ through a fully connected (FC) layer and a gated recurrent unit (GRU), followed by continuous and discrete output heads. In particular, the continuous head parameterizes Gaussian distributions for the movement direction and speed using a state-dependent mean and a learnable, state-independent standard deviation. Additionally, the discrete head produces categorical distributions over the feasible SF and TP options. Accordingly, the joint policy probability used during training is obtained by combining the Gaussian density of the mobility components with the categorical probabilities of the SF and TP selections~\cite{fan2019hybrid}. Moreover, the actor network is optimized by maximizing the clipped surrogate objective with entropy regularization~\cite{yu2022surprising}, given by
\begin{align}
\mathcal{L}^{\text{actor}}(\phi) = \frac{1}{BU} \sum_{b=1}^{B} \sum_{u=1}^{U} \bigg[ &\min\left( \Gamma_{b,u}(\phi) \hat{A}_{b,u}, \Gamma^{\text{clip}}_{b,u} \hat{A}_{b,u} \right) \nonumber \\
&+ \varphi \mathcal{H}[\pi_{\phi}(o_{b,u})] \bigg],
\label{eq:actor_loss}
\end{align}
where $\Gamma_{b,u}(\phi)$ is the probability ratio given by
\[
\Gamma_{b,u}(\phi) =
\frac{\pi_{\phi}(a_{b,u}\mid o_{b,u})}
{\pi_{\phi^{\text{old}}}(a_{b,u}\mid o_{b,u})}
\]
and $\Gamma^{\text{clip}}_{b,u}$ is given by
$
\Gamma^{\text{clip}}_{b,u} = 
\begin{cases}
1 - \epsilon, & \text{if } \Gamma_{b,u} < 1 - \epsilon \\
1 + \epsilon, & \text{if } \Gamma_{b,u} > 1 + \epsilon \\
\Gamma_{b,u}, & \text{otherwise.}
\end{cases}
$

Here, $\mathcal{H}[\pi_\phi]$ in Eq.\eqref{eq:actor_loss} denotes the total policy entropy, obtained by summing the Gaussian entropy of the mobility components and the categorical entropies of the SF and TP components, while $\varphi$ is the entropy coefficient. The shared critic is optimized by minimizing the mean squared error between its value estimate and the $n$-step return, and is expressed as
\begin{equation}
\mathcal{L}^{\text{critic}}(\mu) =
\frac{1}{BU}\sum_{b=1}^{B}\sum_{u=1}^{U}
\left(V_\mu(\mathbf{s}_b)-\hat{R}_{u,b}\right)^2.
\label{eq:critic_loss}
\end{equation}
Here, $\hat{R}_{u,b}$ denotes the $n$-step bootstrapped return for agent $u$ in batch sample $b$. This training structure enables decentralized execution while benefiting from the stability of a centralized critic. Consequently, our proposed GLo-MAPPO training algorithm, shown in Algorithm~\ref{alg:glo_mappo_algo}, begins by initializing the policy $ \{\pi_{\phi_u}\}_{u=1}^{U}$ and value networks $V_\mu$. At each timestep, every agent samples from the shared policy using its local observation $o_u[t]$, UAV-ED association is performed using Algorithm~\ref{alg:gain_based_association}, then constraint violations are checked, and penalties are applied to the agents where necessary. Valid actions lead to environment transitions. The resulting transitions are stored in buffer $\mathcal{D}$. After every $K$ epochs, the advantage function is computed using Eq.~\eqref{eq:advantage_function}, and the actor and critic are updated according to Eqs.~\eqref{eq:actor_loss} and \eqref{eq:critic_loss}, respectively. The training loop repeats for multiple episodes, with periodic evaluations and hyperparameter adjustments. It is important to note that we decouple the UAV-ED association from the MARL framework to reduce action space complexity significantly. Since the association is primarily driven by signal strength, this separation maintains near-optimal performance while allowing the agents to focus specifically on the complex joint optimization of trajectories and communication parameters.

\subsection{Computational Complexity Analysis}
We analyse the computational complexity of our proposed approach across the two phases of its CTDE scheme, and we characterize how the cost scales with the numbers of UAVs $U$ and EDs $V$ for a fixed hidden-layer dimension $H$ of the actor and critic networks. Here, we denote $e_{\max}$, $T$, and $K$ to be the number of training episodes, the episode horizon, and the number of PPO epochs per update, respectively.

\begin{table}[h!]
    \centering
    \caption{Simulation parameters}
    \label{tab:network_env_parameters}
    \resizebox{0.95\columnwidth}{!}{%
    \begin{tabular}{|c|l|c|}
    \hline
    \textbf{Parameter} & \textbf{Description} & \textbf{Value} \\ \hline
    $f$, $W$                     & Carrier frequency, bandwidth          & 868 MHz, 125 kHz \\ \hline
    $U$, $V$                     & Number of GWs, EDs                    & 2$\sim$10, 10$\sim$400 \\ \hline
    $D_{safe}$, $S_{max}$        & UAV safe distance, max velocity       & 3 m, 30 m/s \\ \hline
    $\mathbf{\Psi}$              & Spreading factor                      & [7, 8, 9, 10, 11, 12] \\ \hline
    $\mathcal{P}$                & Transmit power                        & [2, 5, 8, 11, 14] dBm \\ \hline
    $\beta$                      & Cone total angular width              & $120^{\circ}$ \\ \hline
    $\sigma^{2}$                 & Noise power                           & $-117$ dBm \\ \hline
    $\eta_{\text{LoS}}$, $\eta_{\text{NLoS}}$ & LoS, NLoS path-loss coefficients & 0.1, 21 dB \\ \hline
    $\lambda$, $\vartheta$       & Environment-dependent constants       & 0.43, 4.88 \\ \hline
    $A_1$, $A_2$                 & Rician $K$-factor coefficients        & 1.0, 4.39 \\ \hline
    $\eta_{dc}$                  & LoRa ED duty cycle                    & 1\% \\ \hline
    $q^{\mathrm{co}}$            & Co-SF capture threshold               & 6 dB \\ \hline
    $P_c$                        & ED circuit power                      & 1 mW \\ \hline
    $NF$                         & Receiver noise figure                 & 6 dB \\ \hline
    $L$, CR                      & Payload size, coding rate             & 20 B, 4/5 \\ \hline
    $P_{\max}$                   & UAV propulsion power limit            & 200 W \\ \hline
    \end{tabular}%
    }
\end{table}

\begin{table}[h]
    \centering
    \caption{GLo-MAPPO training parameters}
    \label{tab:hyperparams_mappo_sharing}
    \resizebox{0.95\columnwidth}{!}{%
    \begin{tabular}{|c|l|c|}
    \hline
    \textbf{Parameter} & \textbf{Description} & \textbf{Value} \\ \hline
    $H$, -   & Hidden dimension, RL architecture     & 128, GRU \\ \hline
    -, $\alpha$        & Optimizer, learning rate (linearly decayed) & Adam, $3\!\times\!10^{-3} \rightarrow 3\!\times\!10^{-4}$ \\ \hline
    $T$, $B$           & Max steps per episode, replay buffer size & 500, 10 \\ \hline
    $\gamma$, $\epsilon$, $\varphi$ & Discount, clip range, entropy coefficient & 0.90, 0.3, 0.001 \\ \hline
    -, $K$         & Soft target update, PPO epochs         & 0.002, 4 \\ \hline
    -                  & Seeds                                 & 0, 7, 41, 233, 490, 688, 1024 \\ \hline
    $\omega_{1}, \dots, \omega_{7}$ & Reward weights           & 10, 400, $4\times10^{-7}$, 5, 200, 2, 400 \\ \hline
    $\delta_{\text{cap}}$ & Spread saturation distance          & 250 m \\ \hline
    \end{tabular}%
    }
\end{table}

The training phase comprises three components, that is, the gain-based association, the channel and reception modeling, and the learning networks. First, for the gain-based association, the algorithm scans, for each ED $v$, the $U$ UAVs to test the range and capacity constraints and selects the admissible UAV of highest channel gain, costing $O(U)$ per ED and $O(UV)$ over all EDs. Second, the channel and reception modeling forms the block-fading gain matrix at cost $O(UV)$ and evaluates the LoRa capture model over the in-range co-SF and inter-SF interferers of each served ED, costing $O(UV+V^{2})$ per step. Lastly, for the learning networks, each agent observes $o_u=O(V)$ features and a factorized action $a_u = O(\Lambda_{\max})=O(V/U)$, since by construction $\Lambda_{\max}=\lceil V/U \rceil$ and the SF and TP decisions are allocated per associated ED up to this per-UAV capacity. Therefore, one shared-actor pass costs $C_{\mathrm{actor}}=O(VH)$, while the centralized critic maps the global state $s=O(UV)$ to a per-agent value and is evaluated $U$ times, giving $U\,C_{\mathrm{critic}}=O(U^{2}VH)$. These three components enter the total cost differently. The association and the channel and reception modeling are evaluated once per environment step, whereas the learning networks are evaluated over $K$ PPO epochs per update. Accordingly, the channel and reception modeling contributes $O(e_{\max}T\,V^{2})$, while the learning networks contribute $O(e_{\max}TK\,U^{2}VH)$. The overall training cost is therefore $O(e_{\max}TK\,U^{2}VH)$.

In the execution phase, the critic network is no longer required. Each UAV simply maps its local observation to an action through a single actor pass costing $O(VH)$, independent of the rest of the fleet, and $O(UVH)$ system-wide. Overall, the training cost scales quadratically with the number of UAVs $U$ and linearly with the number of EDs $V$. The quadratic dependence on $U$ arises because the centralized critic is evaluated once per agent over a global state of size $O(UV)$. At execution, this quadratic term disappears, and the per-UAV cost becomes independent of $U$, so the system-wide cost grows only linearly in both $U$ and $V$. This analysis is summarized in Table~\ref{tab:complexity}.

\begin{table}[h]
\centering
\tiny
\caption{Computational complexity of GLo-MAPPO.}
\label{tab:complexity}
\footnotesize
\renewcommand{\arraystretch}{1.1}
\setlength{\tabcolsep}{5pt}
\begin{tabular}{llc}
\hline
\textbf{Phase} & \textbf{Component} & \textbf{Cost} \\
\hline
\multirow{3}{*}{Training}
 & Gain-based association (per step)      & $O(UV)$ \\
 & Channel \& reception (per step)        & $O(UV{+}V^{2})$ \\
 & Learning (total)                       & $O(e_{\max}TKU^{2}VH)$ \\
\hline
\multirow{2}{*}{Execution}
 & Per-UAV inference (per step)           & $O(VH)$ \\
 & System-wide inference (per step)       & $O(UVH)$ \\
\hline
\end{tabular}
\end{table}

\vspace{-0.3cm}
\section{Simulation Results}\label{Section_7}

\subsection{Simulation Setup}
To evaluate our proposed approach, we consider a varying number of EDs within a $1$~km~$\times$~$1$~km area. We also deploy a varying number of flying GWs, which travel from their initial positions to a final charging station at $\mathbf{q}_{cs} = [900, 900, 150]$~m while maintaining a fixed altitude of $H = 150$~m. The UAVs are initialized at fixed, equally spaced positions along the lower boundary of the target area, while the ED positions are drawn once from a uniform random distribution over the target area. The total mission duration is $T_{\text{total}}=250$~s, discretized into $T=500$ steps of $\Delta t=0.5$~s.

We implemented the proposed GLo-MAPPO algorithm in PyTorch and integrated it into the EPyMARL framework~\cite{papoudakis2021benchmarking} through a custom MARL environment that follows the OpenAI Gym interface~\cite{brockman2016openai}. Closed-loop validation was performed using the FLoRa framework~\cite{slabicki2018adaptive}, a packet-level simulator built on OMNeT++ and INET. Both MARL training and closed-loop evaluations were executed on a CPU with two Intel Xeon Platinum 8276 processors operating at $2.20$~GHz, providing $56$ CPU cores in total and $376$~GB of RAM. Each MARL training run was allocated $20$ CPU cores, with training results averaged across seven random seeds, while closed-loop evaluation used $2$ CPU cores. Tables~\ref{tab:network_env_parameters} and~\ref{tab:hyperparams_mappo_sharing} detail the communication parameters and DRL hyperparameters, respectively.

Furthermore, we benchmark our proposed approach against multi-agent advantage actor-critic (MAA2C)~\cite{papoudakis2021benchmarking}, counterfactual multi-agent policy gradients (COMA)~\cite{foerster2018counterfactual}, value decomposition networks (VDN)~\cite{sunehag2017value}, monotonic value function factorisation for deep multi-agent reinforcement learning, popularly known as QMIX~\cite{rashid2020monotonic}, and independent proximal policy optimization (IPPO)~\cite{de2020independent}, covering both on-policy and off-policy state-of-the-art MARL algorithms. Additional ablation studies are conducted to evaluate the effects of the optimization parameters and the association scheme.

\subsection{Hyperparameter Sensitivity} \label{hyperparameter_analysis}
\begin{figure}[t!]
\centering
\setlength{\tabcolsep}{2.5pt}
\begin{tabular}{@{}cc@{}}
    \includegraphics[width=0.48\columnwidth]{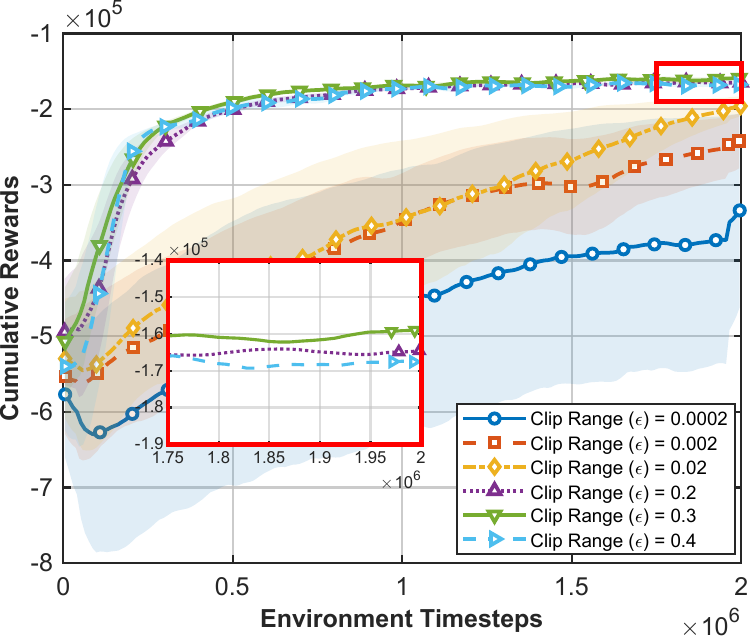} &
    \includegraphics[width=0.48\columnwidth]{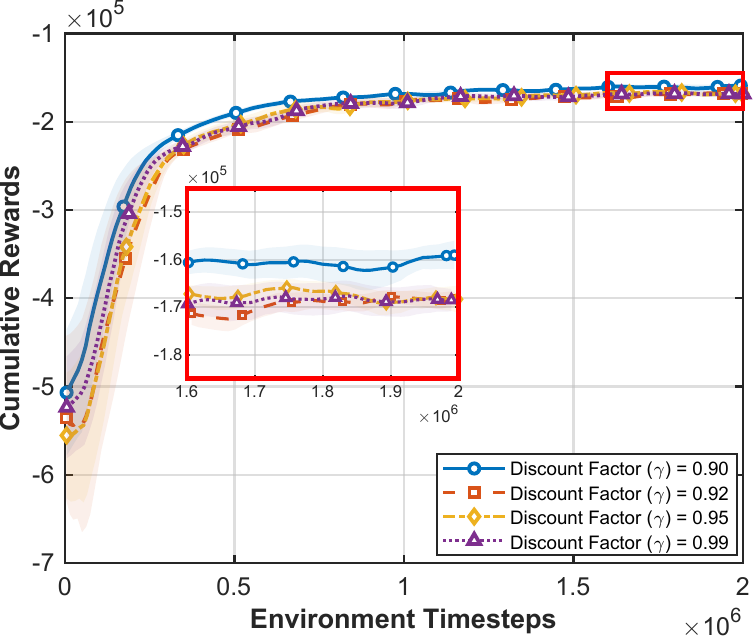} \\
    \small (a) & \small (b)
\end{tabular}
\vspace{-5pt}
\caption{Sensitivity of GLo-MAPPO to (a) the clip range $\epsilon$ and (b) the discount factor $\gamma$.}
\label{fig:training_hyperparameters}
\vspace{-14pt}
\end{figure}

Fig.~\ref{fig:training_hyperparameters} examines the hyperparameter sensitivity of our proposed GLo-MAPPO approach with $2$ GWs and $50$ deployed LoRa EDs for clip ranges of $\epsilon \in \{0.0002,0.002,0.02, 0.2,0.3,0.4\}$ and discount factor $\gamma \in\{0.9, 0.92, 0.95, 0.99\}$. Specifically, Fig.~\ref{fig:training_hyperparameters}(a) shows that moderate clip ranges of $\epsilon\in\{0.2,0.3,0.4\}$ converge faster and achieve higher cumulative rewards than the smaller values. In particular, $\epsilon=0.3$ provides the highest cumulative reward, indicating a suitable balance between restricting abrupt policy changes and allowing sufficient policy improvement. In contrast, $\epsilon\in\{0.0002,0.002,0.02\}$ overly constrains the updates, resulting in slower learning and lower final rewards. Next, Fig.~\ref{fig:training_hyperparameters}(b) evaluates the discount factor $\gamma$. Although all examined values eventually stabilize, $\gamma=0.9$ converges faster and attains the highest final cumulative reward. Increasing $\gamma$ places greater emphasis on distant rewards but provides no performance advantage in the considered environment, indicating that a short- to medium-term planning horizon is more effective. Accordingly, we adopt $\epsilon=0.3$, and $\gamma=0.9$ for the subsequent simulations. Therefore, Table~\ref{tab:hyperparams_mappo_sharing} provides the detailed DRL parameters.

\begin{figure*}[t]
\centering
\setlength{\tabcolsep}{2pt}
\begin{tabular}{@{}ccc@{}} 
    \includegraphics[width=0.28\textwidth]{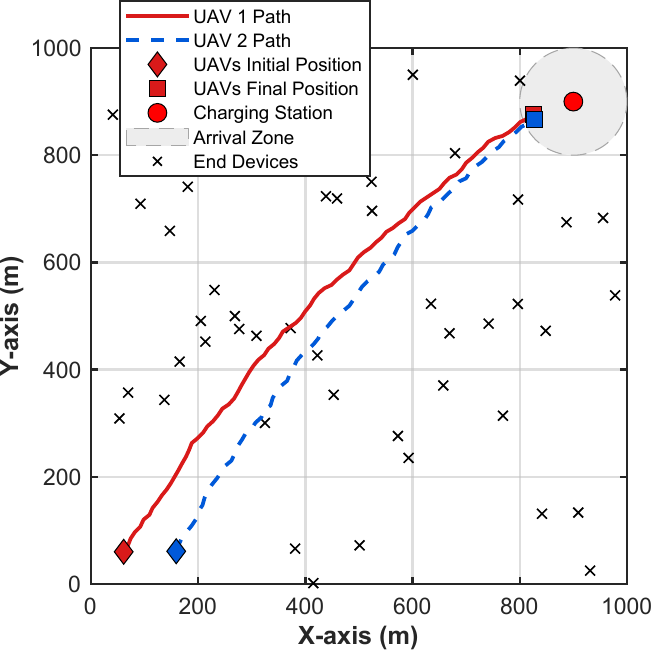} &
    \includegraphics[width=0.35\textwidth]{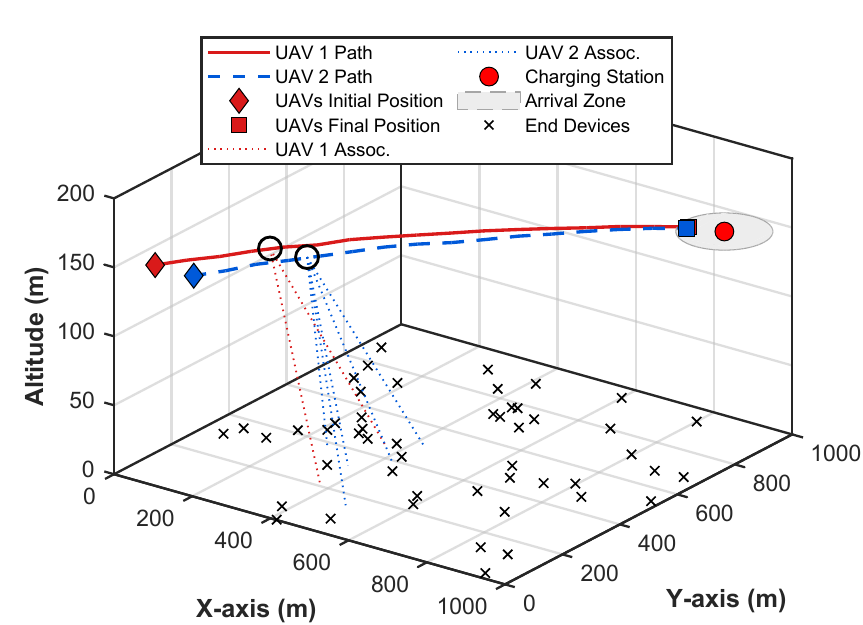} &
    \includegraphics[width=0.3\textwidth]{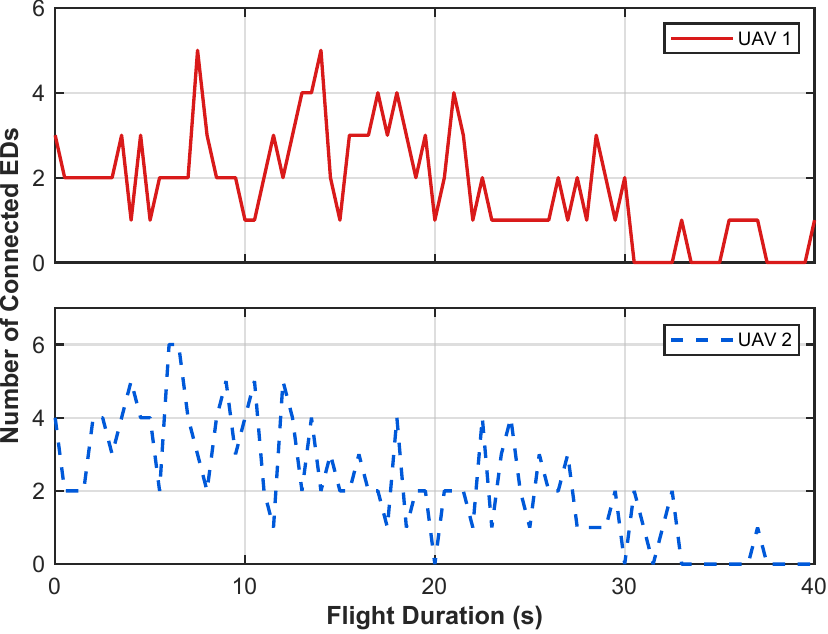} \\
    \small (a) &
    \small (b) &
    \small (c)
\end{tabular}
\setlength{\tabcolsep}{6pt} 
\vspace{-5pt} 
\caption{(a) Optimized 2D trajectories, (b) Optimized 3D trajectory with ED-UAV association, (c) Number of connected EDs by each UAV over time.}
\label{fig:UAV_performance_comparison}
\vspace{-13pt}
\end{figure*}

\begin{figure*}[t]
\centering
\setlength{\tabcolsep}{2pt}
\begin{tabular}{@{}ccc@{}} 
    \includegraphics[width=0.29\textwidth]{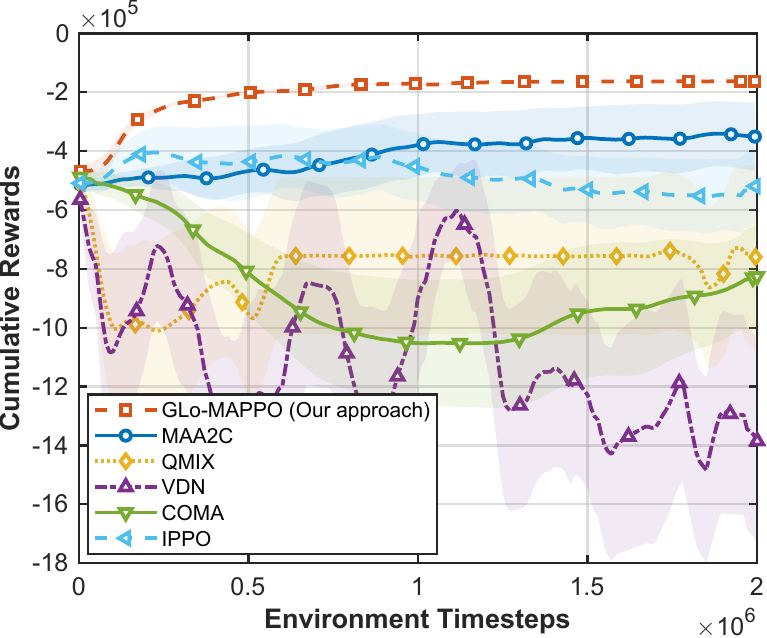} &
    \includegraphics[width=0.28\textwidth]{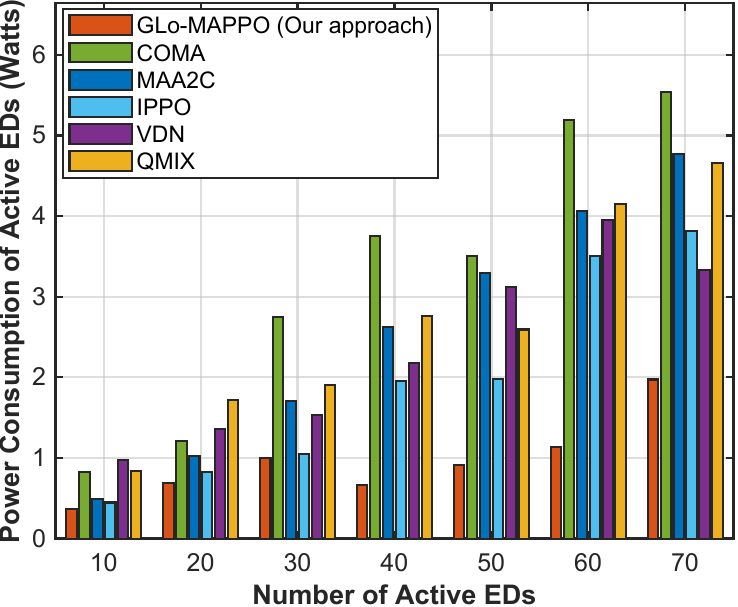} &
    \includegraphics[width=0.28\textwidth]{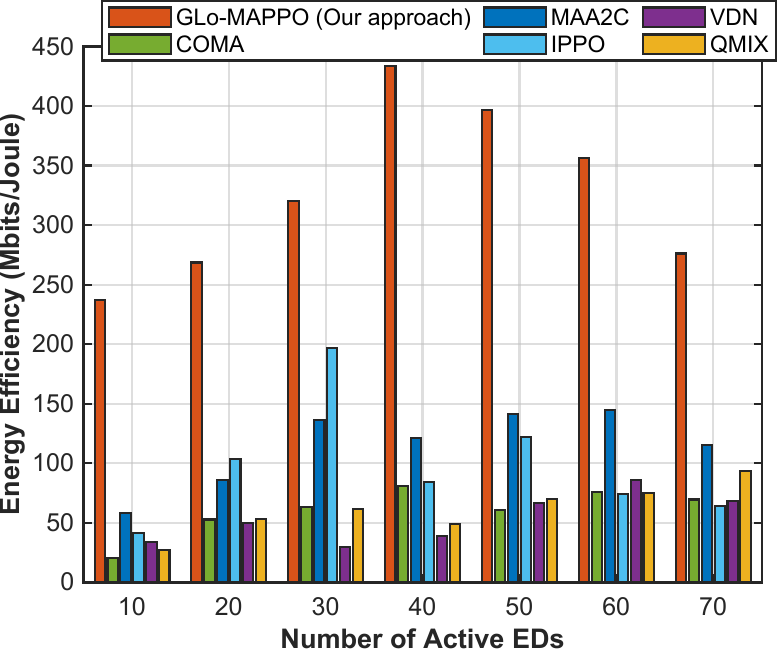} \\
    \small (a) &
    \small (b) &
    \small (c)
\end{tabular}
\setlength{\tabcolsep}{6pt}
\vspace{-5pt} 
\caption{(a) Cumulative rewards, (b) Power consumption of Active EDs, (c) System Energy efficiency.}
\label{fig:rewards_power_ee}
\vspace{-10pt} 
\end{figure*}

\subsection{Optimized UAV Trajectories and Association Dynamics} \label{uav_trajectories_with_association}
Fig.~\ref{fig:UAV_performance_comparison} shows a scenario with $2$ UAVs serving $50$ EDs. Specifically, Fig.~\ref{fig:UAV_performance_comparison}(a) shows the 2D flight paths of the two UAVs. Both start from their initial positions and steer toward the denser groups of EDs, so as to serve as many devices as possible along the way. The two paths never cross and stay clearly apart for the whole mission, which confirms that the safety distance constraint is respected and collisions are avoided. By the end of the flight, both UAVs reach their arrival zones. Fig.~\ref{fig:UAV_performance_comparison}(b) gives a 3D view of the same mission, with the active UAV-ED links shown at time $t=20$s. Each UAV keeps serving the ground EDs below it while flying toward the charging station, and the association links at $t=18$~s illustrates how the associations change as the UAVs move. Fig.~\ref{fig:UAV_performance_comparison}(c) presents the number of connected EDs for each UAV over the flight. Both UAVs serve several EDs during the early and middle part of the mission, when they fly over the populated area. As the UAVs approach their arrival zones, they move away from the EDs, so the number of connected devices falls and reaches zero once they land.

\subsection{Performance Analysis of the Proposed GLo-MAPPO} \label{ee_performance_comparison}
Fig.~\ref{fig:rewards_power_ee} compares GLo-MAPPO against state-of-the-art MARL benchmarks. In particular, Fig.~\ref{fig:rewards_power_ee}(a) shows the cumulative rewards over environment timesteps for a setup with $2$~GWs and $70$~EDs. In particular, our proposed GLo-MAPPO reaches the highest reward, converges quickly, and stays stable thereafter. It can also be observed that VDN, QMIX, and COMA are the weakest, with much lower and more volatile rewards, while the policy-based methods, such as MAA2C and IPPO, converge more smoothly, but their rewards still stay well below our approach. In Fig.~\ref{fig:rewards_power_ee}(b), we show the total uplink EDs power consumption as the number of active EDs grows from $10$ to $70$. Across the evaluated deployed EDs, our proposed GLo-MAPPO uses the least power at every density. Conversely, other benchmarks such as COMA, MAA2C, IPPO, VDN and QMIX consume the most, especially as the network grows denser. Similarly, Fig.~\ref{fig:rewards_power_ee}(c) presents the system energy efficiency across varying EDs. As expected, our proposed GLo-MAPPO approach keeps the highest energy efficiency throughout and clearly outperforms every benchmark at all densities.

\subsection{Scalability Study} \label{scalability_study}
\begin{figure}[t!]
\centering
\setlength{\tabcolsep}{2.5pt}
\begin{tabular}{@{}cc@{}} 
    \includegraphics[width=0.45\columnwidth]{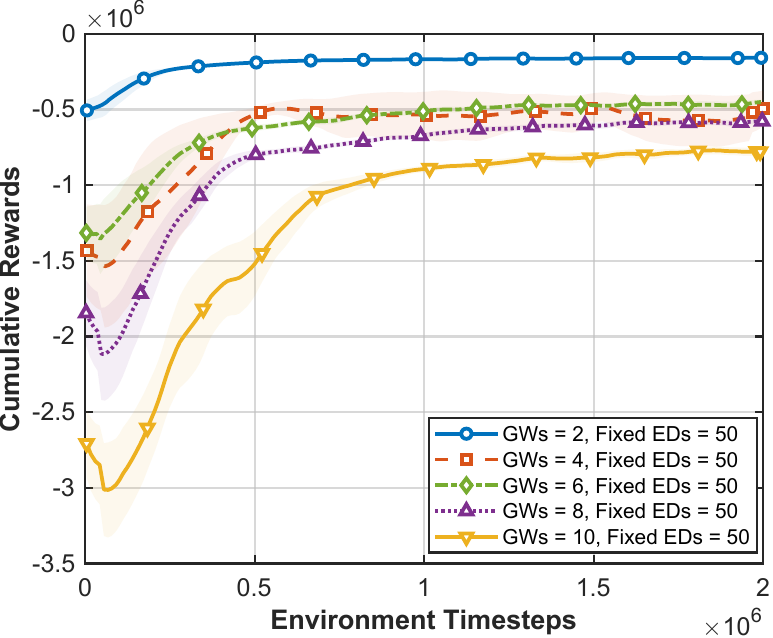} &
    \includegraphics[width=0.45\columnwidth]{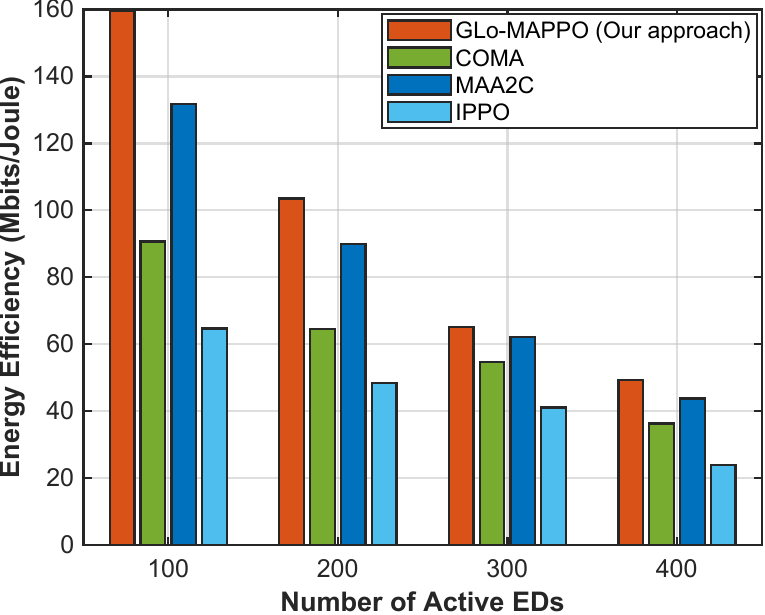} \\
    \small (a) &
    \small (b)
\end{tabular}
\vspace{-5pt} 
\caption{Scalability Analysis of system energy efficiency (a) Training convergence under varying numbers of UAVs, and (b) Large-scale numbers of active EDs.}
\label{fig_scalability_study}
\vspace{-14pt}
\end{figure}

In this subsection, we assess scalability with regard to the varying number of UAVs as well as the number of EDs. First, the effect of varying the number of UAVs from $U=2$ to $10$, with a fixed number of $V=50$ EDs, is shown in Fig.~\ref{fig_scalability_study}(a). As shown in the figure, all five configurations learn and settle at stable reward values, indicating that our approach continues to converge as the fleet grows. The reward is lower for larger UAV fleets because of the increased complexity of the environment, including collision, spreading-distance, and arrival penalties, along with a more challenging coordination problem. Next, we fix the number of UAVs at $2$ and vary the number of active EDs from $100$ to $400$, comparing our approach against the three strongest baselines. As shown in Fig.~\ref{fig_scalability_study}(b), our approach achieves the highest EE at every density. This shows our proposed GLo-MAPPO remains superior even as the number of active EDs grows.

Additionally, Table~\ref{tab:ed_density_metrics} compares GLo-MAPPO with the three strongest baselines across small-scale ($50$, $60$ and $70$ EDs) and large-scale ($200$ and $400$ EDs) deployments. As can be seen, our proposed GLo-MAPPO achieves the best result in nearly all configurations. Moreover, as we averaged across all densities, GLo-MAPPO improves energy efficiency by $96.0\%$ over the best baseline and $168.1\%$ over the baseline average. It also reduces ED transmit power by $34.8\%$ and $42.9\%$, and UAV propulsion power by $15.3\%$ and $59.8\%$, respectively. Moreover, its propulsion power remains stable between $30.57$ and $31.71$~kW, while the baselines vary considerably, and it records zero UAV collisions at every density. Hence, these show the reliability and scalability of GLo-MAPPO for dense IoT deployments.

\subsection{Ablation Study} \label{ablation_study}
\begin{figure}[t]
\centering
\setlength{\tabcolsep}{2.5pt}
\begin{tabular}{@{}cc@{}} 
    \includegraphics[width=0.46\columnwidth]{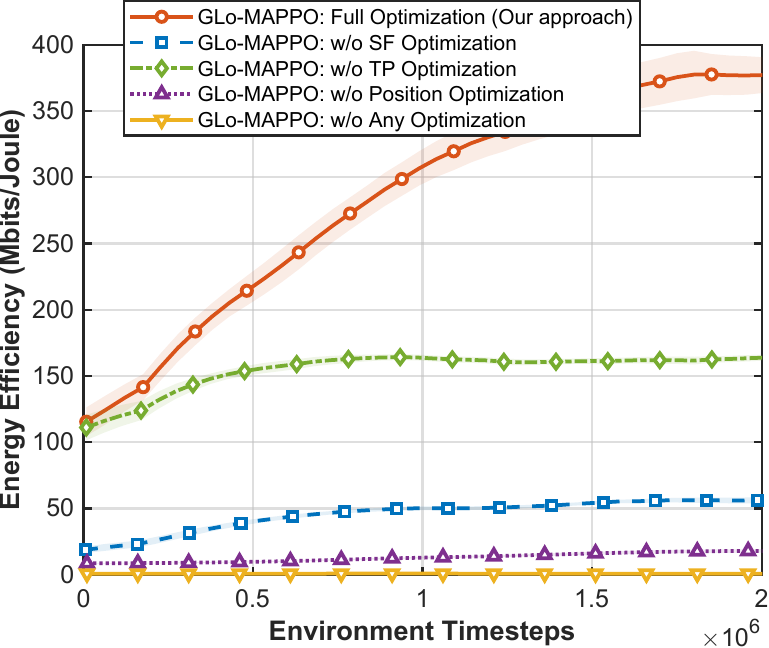} &
    \includegraphics[width=0.46\columnwidth]{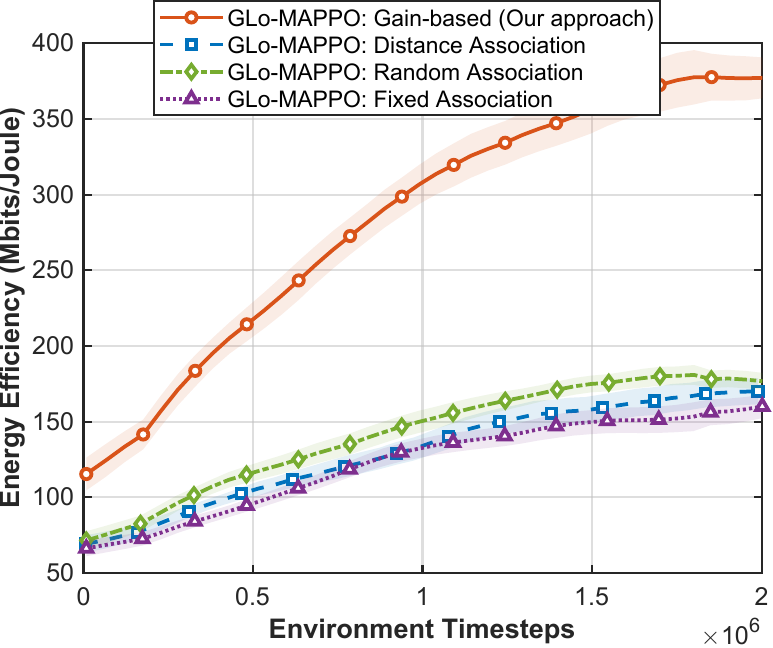} \\
    \small (a) &
    \small (b)
\end{tabular}
\vspace{-5pt} 
\caption{Ablation study of our GLo-MAPPO approach: (a) Optimization parameters, and (b) Association schemes.}
\label{fig_ablation_study}
\vspace{-14pt}
\end{figure}

Fig.~\ref{fig_ablation_study} presents our ablation study for the joint optimization parameters and association scheme. In particular, the ablation studies were conducted with a configuration of $2$~UAVs and $50$~EDs. In Fig.~\ref{fig_ablation_study}(a), we examine four configurations: (i) a fixed spreading factor of $12$ (GLo-MAPPO w/o SF optimization), (ii) a fixed transmission power of $TP=8$~dBm (GLo-MAPPO w/o TP optimization), (iii) static UAV positioning (GLo-MAPPO w/o position optimization), and (iv) a non-optimized baseline where all three parameters remain fixed (GLo-MAPPO w/o any optimization). It can be observed in Fig.~\ref{fig_ablation_study}(a) that our full GLo-MAPPO framework achieves the highest energy efficiency, whereas having a static position causes the largest performance loss, followed by only SF and TP optimization. When all three variables are fixed, the performance remains near zero. These results confirm that all optimization components contribute to the proposed framework, with UAV position optimization having the greatest impact. Fig.~\ref{fig_ablation_study}(b) further compares the proposed gain-based ED association with distance-based, random, and fixed association schemes. As expected, our proposed gain-based scheme outperforms all three baselines, while random association performs best among the baselines and fixed association performs worst. 

\begin{figure*}[t]
\centering
\setlength{\tabcolsep}{2pt}
\begin{tabular}{@{}ccc@{}}
    \includegraphics[width=0.3\textwidth]{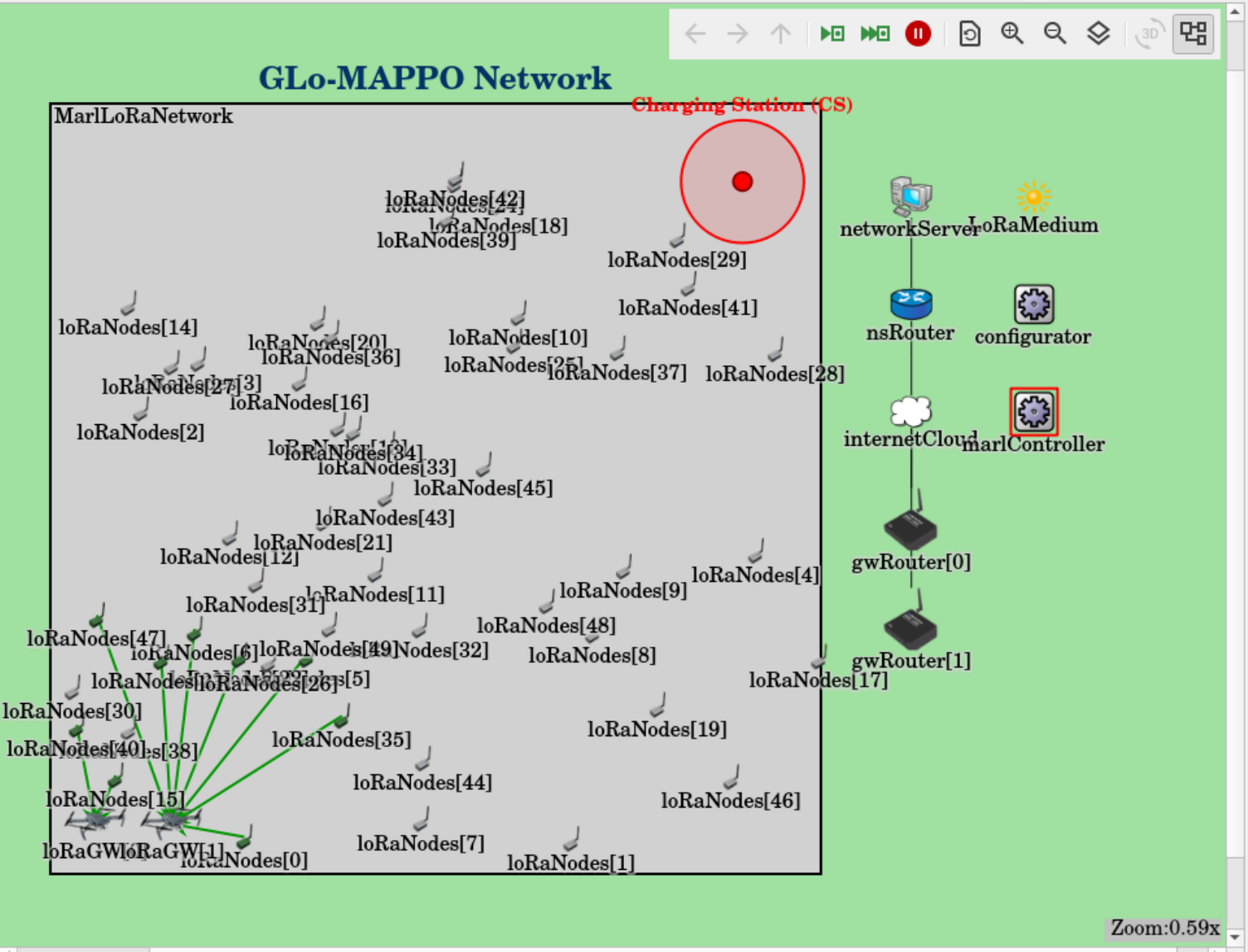} &
    \includegraphics[width=0.3\textwidth]{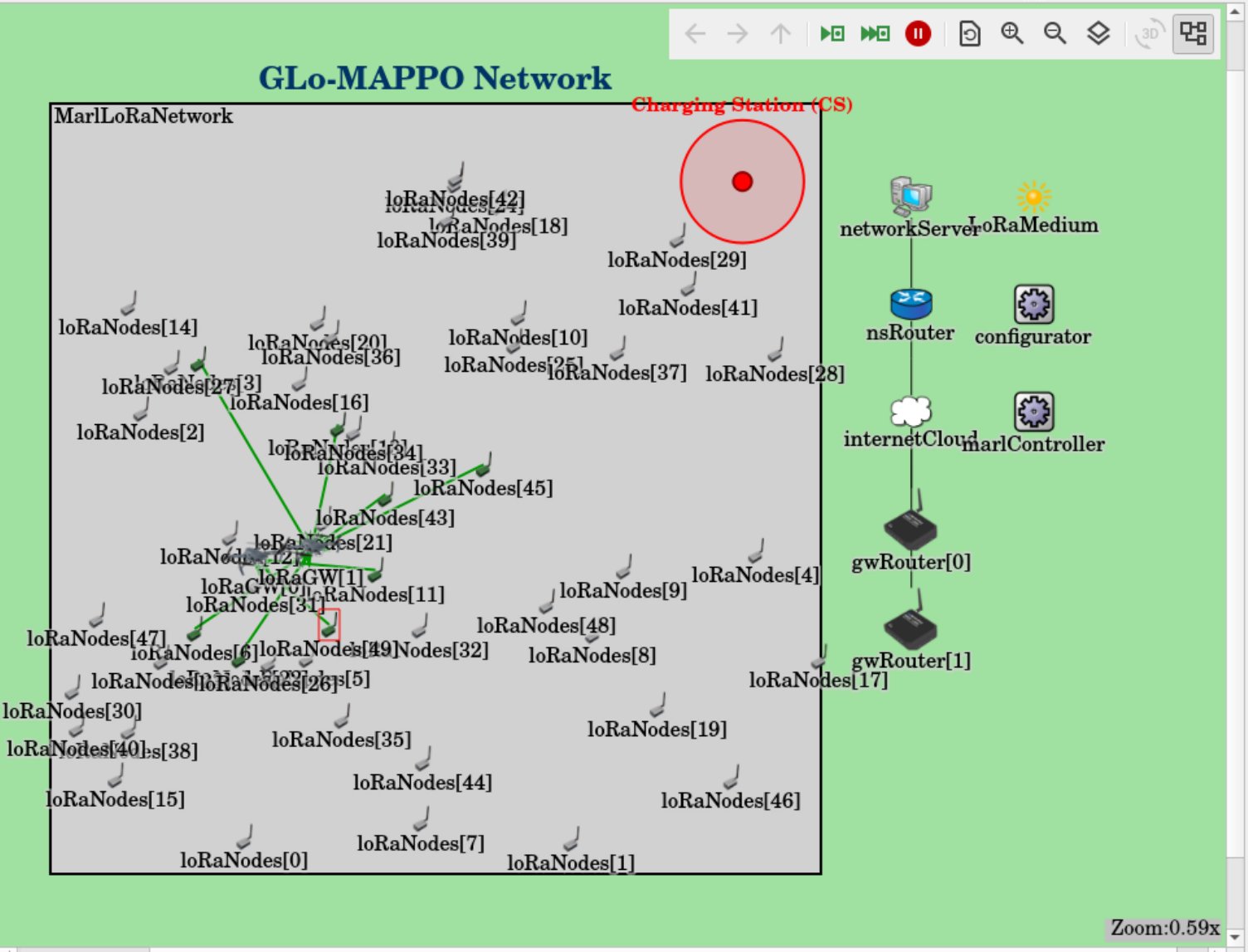} &
    \includegraphics[width=0.3\textwidth]{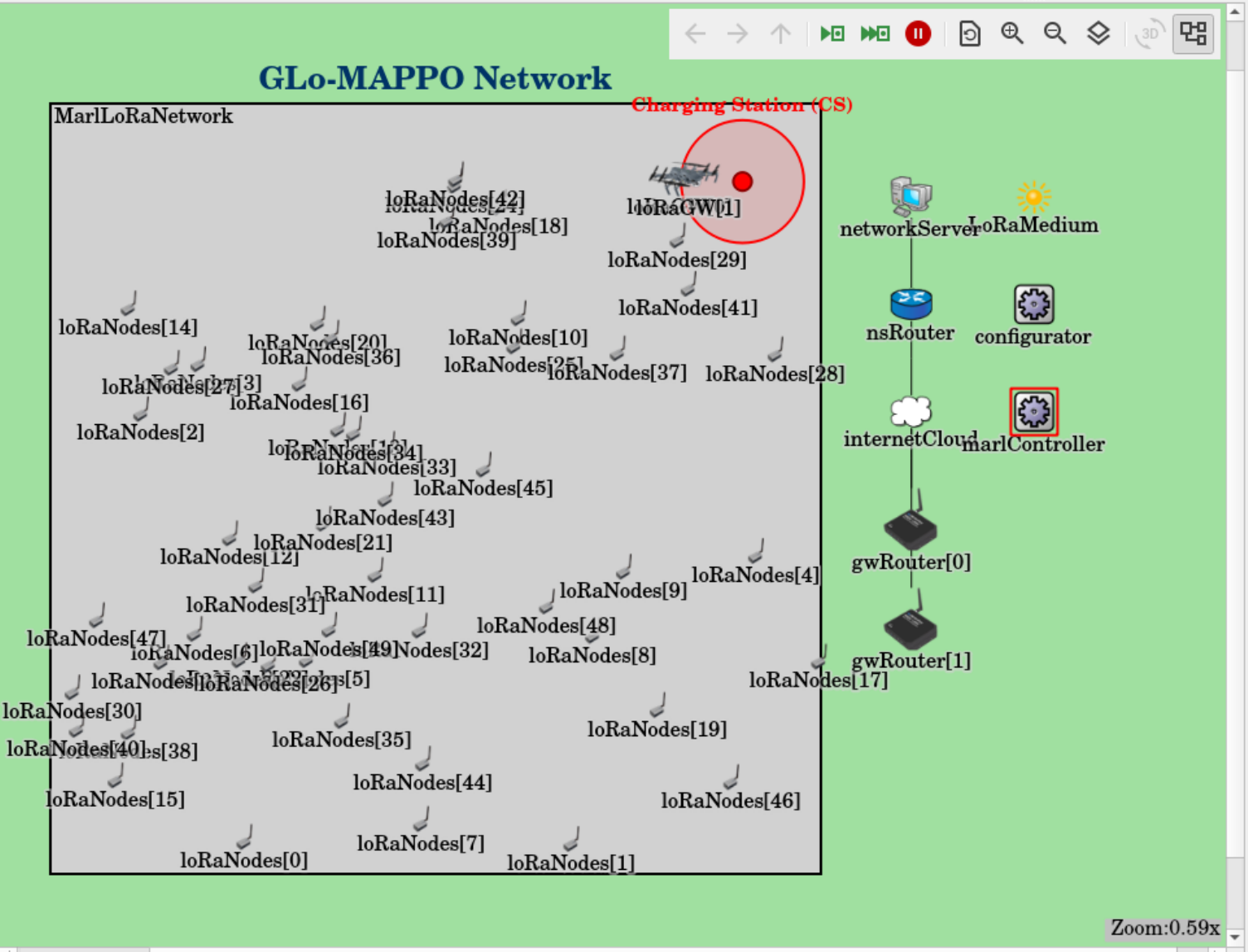} \\
    \small (a) &
    \small (b) &
    \small (c) 
\end{tabular}
\setlength{\tabcolsep}{6pt}
\vspace{-5pt}
\caption{Closed-loop co-simulation in OMNeT++/FLoRa: (a) Initial deployment, (b)  Mid-mission, and (c) Arrival at the destination.}
\label{fig:flora_mission}
\vspace{-10pt}
\end{figure*}

\subsection{Validation via Packet-Level LoRa Co-Simulation with OMNeT++/FLoRa}\label{sec:cosim}

To further validate our proposed GLo-MAPPO framework beyond the analytical environment used for training, we evaluate the trained policy\footnote{We evaluate the model obtained with random seed $41$ at its final checkpoint.} in a closed loop with FLoRa~\cite{slabicki2018adaptive}, a packet-level LoRa framework built on OMNeT++~6.0.3 and INET~4.4.1. In this setting, the learned policy no longer acts on an abstract rollout, but rather drives real LoRa gateways whose mobility, radio configuration, and uplink transmissions are executed step-by-step by the OMNeT++/INET discrete-event stack. Communication between OMNeT++/FLoRa and the Python back-end is established over TCP/IP sockets (BSD/POSIX sockets on Linux), forming a lock-step, request/response loop that ensures deterministic synchronization between the simulator and the policy. At every simulation timestep, the FLoRa control interface transmits the current network measurements to the Python back-end and blocks until it receives the policy's decision. Accordingly, each control step comprises three stages, given as follows:
\begin{itemize}
    \item FLoRa transmits, for every UAV-ED pair, the measured channel gain $G_{u,v}[t]$ and a freshness flag indicating whether the ED lies within $R_{comm}$, from which the agent observation of Eq.~\eqref{eq_agent_observation} is assembled.
    \item The GLo-MAPPO trained policy returns, for each gateway, its movement decision and the per-ED SF and TP allocations $\boldsymbol{\psi}_u[t]$ and $\mathbf{p}_u[t]$.
    \item Lastly, FLoRa applies these decisions by repositioning the gateways, configuring each ED's LoRa radio, and gating the served transmissions.
\end{itemize}

Consequently, the LoRa capture, co-/inter-SF interference given in Table~\ref{tab:thresholds}, the duty-cycle, and association logic are retained in the loop and driven by the FLoRa-measured gain, so the active channel model directly governs reception.

Additionally, to assess both integration reliability and robustness to the channel model, the policy is fed three channel tiers of increasing realism, each benchmarked against our GLo-MAPPO standalone analytic environment. First, Tier-1 ports the same analytic G2A model given in Eq.~\eqref{eqt_G2A_los} into the FLoRa framework as an integration check. Next, Tier-2 augments it with independent log-normal shadowing of $\sigma = 8$~dB as a robustness test, and lastly, Tier-3 substitutes FLoRa's native measurement-based Oulu path-loss model~\cite{petajajarvi2015coverage}. We perform the evaluation with $U = 2$ gateways and $V = 50$ EDs over $T = 500$ control steps, and we average over $10$ episodes per tier. Throughout the flight mission, the gateways autonomously adapt their trajectories and per-ED allocations to the evolving channel and converge to the CS by the episode's end. Lastly, Fig.~\ref{fig:flora_mission} illustrates the validated flight mission, similar to the GLo-MAPPO standalone training in Fig.~\ref{fig:UAV_performance_comparison}, where the active UAV-ED service associations are shown in green. A video demonstration is also available.\footnote{Demonstration video: \url{https://youtu.be/5H73WmQZgK4}} 

Furthermore, we report the closed-loop metrics from the retained LoRa reception model operating on the FLoRa-measured gain of the active tier. In particular, the packet delivery ratio (PDR) is the fraction of attempted ED uplinks successfully received, where an uplink counts as delivered only if it clears the reception threshold and survives the co-/inter-SF capture given in Table~\ref{tab:thresholds} and the LoRa duty-cycle scheme. The coverage is the percentage of the $V$ EDs achieving at least one successful reception, and $EE_{\mathrm{sys}}$ is as defined in Section~\ref{ee_model}. Here, the standalone reference is the single-seed 41 checkpoint used to drive the co-simulation. As shown in Table~\ref{tab:cosim}, Tier-1 reproduces the GLo-MAPPO standalone metrics to within $0.29\%$. Under Tier-2, coverage and PDR remain unchanged while $EE_{\mathrm{sys}}$ deviates by only $0.38\%$, and this shows robustness to realistic channel randomness. Finally, under Tier-3, the Oulu model is a terrestrial path-loss model and is not intended for a 150~m ground-to-air link, so its large $EE_{\text{sys}}$ deviation of about $79\%$, which is expected. Delivery and coverage, which depend on LoRa's low reception thresholds, stay almost unchanged. This shows that reception is robust across channel models, while the achievable rate is sensitive to the propagation assumption.

\begin{table*}[!t]
\centering
\footnotesize
\caption{Performance of GLo-MAPPO against baselines across five ED densities.}
\label{tab:ed_density_metrics}
\renewcommand{\arraystretch}{1.2}
\setlength{\tabcolsep}{4pt}
\begin{tabular}{|l|l|c|c|c|c|c|c|c|}
\hline
\multirow{2}{*}{Metric} & \multirow{2}{*}{Algorithm} & \multicolumn{3}{c|}{Small-scale} & \multicolumn{2}{c|}{Large-scale} & \multicolumn{2}{c|}{Improvement (\%)} \\
\cline{3-5}\cline{6-7}\cline{8-9}
 & & 50 EDs & 60 EDs & 70 EDs & 200 EDs & 400 EDs & vs best baseline & vs avg. all baselines \\
\hline
\multirow{4}{*}{EE (Mbit/J) $\uparrow$} & GLo-MAPPO (Ours) & \textbf{397.59} & \textbf{355.64} & \textbf{275.91} & \textbf{91.87} & \textbf{49.39} & \textbf{+96.0} & \textbf{+168.1} \\
 & COMA & 60.10 & 75.95 & 69.41 & 64.27 & 36.27 &  &  \\
 & MAA2C & 141.67 & 144.68 & 115.65 & 89.79 & 43.77 &  &  \\
 & IPPO & 121.88 & 74.56 & 64.33 & 48.12 & 23.95 &  &  \\
\hline
\noalign{\vskip 3pt}
\hline
\multirow{4}{*}{ED Tx power (W) $\downarrow$} & GLo-MAPPO (Ours) & \textbf{0.91} & \textbf{1.14} & \textbf{1.99} & 11.25 & \textbf{22.61} & \textbf{+34.8} & \textbf{+42.9} \\
 & COMA & 3.49 & 5.19 & 5.54 & 12.73 & 24.98 &  &  \\
 & MAA2C & 3.29 & 4.06 & 4.76 & 12.08 & 24.45 &  &  \\
 & IPPO & 1.98 & 3.51 & 3.83 & \textbf{10.93} & 25.04 &  &  \\
\hline
\noalign{\vskip 3pt}
\hline
\multirow{4}{*}{\shortstack{UAVs' propulsion\\power (kW) $\downarrow$}} & GLo-MAPPO (Ours) & \textbf{30.57} & \textbf{30.88} & \textbf{30.74} & \textbf{31.06} & \textbf{31.71} & \textbf{+15.3} & \textbf{+59.8} \\
 & COMA & 122.82 & 97.15 & 80.57 & 68.73 & 72.90 &  &  \\
 & MAA2C & 41.16 & 31.78 & 56.23 & 31.35 & 32.28 &  &  \\
 & IPPO & 94.83 & 134.24 & 122.20 & 85.00 & 108.72 &  &  \\
\hline
\noalign{\vskip 3pt}
\hline
\multirow{4}{*}{UAV collisions $\downarrow$} & GLo-MAPPO (Ours) & \textbf{0.00} & \textbf{0.00} & \textbf{0.00} & \textbf{0.00} & \textbf{0.00} & \textbf{+100.0} & \textbf{+100.0} \\
 & COMA & 4.86 & 5.71 & 3.14 & 6.86 & 2.00 &  &  \\
 & MAA2C & 5.14 & 3.57 & 7.43 & 0.67 & 0.10 &  &  \\
 & IPPO & 6.57 & 16.86 & 10.86 & 11.52 & 15.33 &  &  \\
\hline
\end{tabular}
\end{table*}

\section{Discussion and Future Work}\label{Section_8}
The results presented in this study show that our proposed GLo-MAPPO effectively addresses the joint optimization of spreading factor selection, transmission power allocation, UAV trajectories, and ED association in multi-UAV LoRa networks under partial observability. Despite these promising results, several aspects deserve further discussion and motivate future research directions. For instance, the proposed framework decouples ED-UAV association from the agents' action space through a gain-based scheme. As discussed in Section~\ref{section_glo_mappo_algo_B}, this design substantially reduces the action-space complexity while allowing the agents to focus on the joint optimization of trajectories and resource allocation. Nevertheless, fully integrating association decisions into the learned policy may unlock additional coordination gains, particularly when capacity quotas and interference coupling become more influential. A promising direction is to incorporate association decisions into heterogeneous multi-agent transformer-based (MAT) architectures, which could capture inter-agent dependencies more expressively while managing the complexity.

Next, although our framework handles realistic channel randomness well, it does not yet account for hardware failures that may arise in practical deployments, such as motor or battery degradation, transceiver faults, or a complete UAV outage. From a deployment perspective, such failures could be addressed by coupling the framework with an onboard fault-detection and status-reporting mechanism, allowing the healthy UAVs to re-plan their trajectories and re-associate the affected EDs through the existing gain-based scheme, while the impaired UAV performs a safe return-to-base procedure. Realizing such fault tolerance would further require incorporating hardware reliability models into the deployment pipeline and validating the system on a physical testbed under induced fault conditions, which we leave for future work.

\begin{table}[h!]
\centering
\tiny
\caption{Closed-loop co-simulation across three channel tiers versus the standalone analytic environment.}
\label{tab:cosim}
\renewcommand{\arraystretch}{1.2}
\setlength{\tabcolsep}{4pt}
\resizebox{\columnwidth}{!}{%
\begin{tabular}{|l|c|c|c|}
\hline
Configuration & PDR (\%) & $EE_{\mathrm{sys}}$ (Mbit/J) & Coverage (\%) \\
\hline
GLo-MAPPO (Standalone) & 83.89 & 346.6 & 82.00 \\
Tier-1 (G2A)           & 83.89 & 345.6 & 82.00 \\
Tier-2 (G2A+$\sigma$)  & 83.89 & 345.3 & 82.00 \\
Tier-3 (Oulu)          & 82.15 & 72.6 & 82.00 \\
\hline
\end{tabular}}
\end{table}

\section{Conclusion}\label{Section_9}
In this work, we studied an uplink data-collection scheme for multi-UAV LoRa networks by formulating the joint optimization of spreading factors, transmission powers, UAV trajectories, and ED associations as a weighted system energy-efficiency maximization problem. We transformed this problem into a POSG and proposed the GLo-MAPPO framework, which leverages a CTDE paradigm together with a gain-based association scheme to achieve stable, cooperative learning under partial observability. Simulation results demonstrated that GLo-MAPPO significantly improves training rewards, reduces ED power consumption, enhances overall energy efficiency, and consistently outperforms state-of-the-art MARL benchmarks across varying network densities and scales.

\section*{Acknowledgment}
Jamal Bentahar is funded by Khalifa University of Science and Technology through the Faculty Start Up Grant Program under Project ID: KU-INT-FSU-2024-8471000001-PD\#10261. This research was supported by KU Research Center for Advanced Intelligent Systems (AIS), Khalifa University of Science and Technology (KU-AIS).

The computing resources for this study were provided by the Toubkal High-Performance Computing (HPC) facilities at Mohammed VI Polytechnic University (UM6P), Benguerir, Morocco.

\balance
\bibliographystyle{IEEEtran}
\bibliography{biblio}

@article{jouhari2023survey,
  title={{A survey on scalable LoRaWAN for massive IoT: Recent advances, potentials, and challenges}},
  author={Jouhari, Mohammed and Saeed, Nasir and Alouini, Mohamed-Slim and Amhoud, El Mehdi},
  journal={IEEE Commun. Surveys \& Tutorials},
  year={2023}
}

@ARTICLE{11037543,
  author={Song, Xiaoyu and Chin, Kwan-Wu},
  journal={IEEE Trans. on Vehicular Technology}, 
  title={{Maximizing Uplink and Downlink Transmissions in RSMA Wirelessly Powered IoT Networks}}, 
  year={2025},
  volume={74},
  number={11},
  pages={17652-17665}}

@ARTICLE{11106313,
  author={Ao, Tianyong and Li, Haoqiang and Zhang, Kaixin and Shi, Huaguang and Shi, Lei and Liu, Fuqiang and Zhou, Yi},
  journal={IEEE Trans. on Vehicular Technology}, 
  title={{Heterogeneous UAVs Trajectory Optimization for Post-Disaster Target Search Based on MARL With Graph Attention Network}}, 
  year={2026},
  volume={75},
  number={1},
  pages={1412-1426}}

@inproceedings{orazi2018first ,
  title={{A first step toward an IoT network dedicated to the sustainable development of a territory}},
  author={Orazi, Gilles and Fontaine, Genevieve and Chemla, Philippe and Zhao, Mengxuan and Cousin, Philippe and Le Gall, Franck},
  booktitle={Global Internet of Things Summit},
  year={2018}
}

@article{fakhruldeen2024enhancing,
  title={{Enhancing smart home device identification in WiFi environments for futuristic smart networks-based IoT}},
  author={Fakhruldeen, Hassan Falah and Saadh, Mohamed J and Khan, Samiullah and Salim, Nur Agus and Jhamat, Naveed and Mustafa, Ghulam},
  journal={International J. of Data Science and Analytics},
  year={2024}
}

@ARTICLE{11175095,
  author={Xia, Yang and Zhang, Haixia and Tian, Jie and Yuan, Dongfeng},
  journal={IEEE Trans. on Vehicular Technology}, 
  title={{Dynamic Task Offloading and Resource Allocation for Vehicular Edge Computing Networks Based on Deep Reinforcement Learning}}, 
  year={2026},
  volume={75},
  number={3},
  pages={4977-4986}}

@article{gong2023modeling,
  title={{Modeling power consumptions for multirotor UAVs}},
  author={Gong, Hao and Huang, Baoqi and Jia, Bing and Dai, Hansu},
  journal={IEEE Trans. on Aerospace and Electronic Systems},
  volume={59},
  number={6},
  year={2023},
}

@article{yu2022surprising,
  title={{The surprising effectiveness of PPO in cooperative multi-agent games}},
  author={Yu, Chao and Velu, Akash and Vinitsky, Eugene and Gao, Jiaxuan and Wang, Yu and Bayen, Alexandre and Wu, Yi},
  journal={Advances in Neural Information Processing Systems},
  volume={35},
  year={2022}
}

@ARTICLE{zeng2019energy,
  author={Zeng, Yong and Xu, Jie and Zhang, Rui},
  journal={IEEE Trans. on Wireless Commun.}, 
  title={{Energy Minimization for Wireless Communication With Rotary-Wing UAV}}, 
  year={2019},
  volume={18},
  number={4},
  pages={2329-2345}}

@ARTICLE{9310195,
  author={Shi, Yongpeng and Xia, Yujie and Gao, Ya},
  journal={IEEE Access}, 
  title={{Joint Gateway Selection and Resource Allocation for Cross-Tier Communication in Space-Air-Ground Integrated IoT Networks}}, 
  year={2021},
  volume={9},
  number={},
  pages={4303-4314},
  }

@article{zhang2018fast,
  title={{Fast deployment of UAV networks for optimal wireless coverage}},
  author={Zhang, Xiao and Duan, Lingjie},
  journal={IEEE Trans. on Mobile Computing},
  volume={18},
  number={3},
  pages={588--601},
  year={2018},
  publisher={IEEE}
}

@article{zhou2022joint,
  title={{Joint optimization of mobility and reliability-guaranteed air-to-ground communication for uavs}},
  author={Zhou, Jianshan and others},
  journal={IEEE Trans. on Mobile Computing},
  volume={23},
  number={1},
  pages={566--580},
  year={2022},
  publisher={IEEE}
}

@inproceedings{papoudakis2021benchmarking,
   title = {{Benchmarking Multi-Agent Deep Reinforcement Learning Algorithms in Cooperative Tasks}},
   author = {Georgios Papoudakis and Filippos Christianos and Lukas Schäfer and Stefano V. Albrecht},
   booktitle = {Proceedings of the Neural Information Processing Systems Track on Datasets and Benchmarks},
   year = {2021},
}

@inproceedings{foerster2018counterfactual,
  title={{Counterfactual multi-agent policy gradients}},
  author={Foerster, Jakob and Farquhar, Gregory and Afouras, Triantafyllos and Nardelli, Nantas and Whiteson, Shimon},
  booktitle={Proceedings of the AAAI conference on artificial intelligence},
  volume={32},
  number={1},
  year={2018}
}

@article{sunehag2017value,
  title={{Value-decomposition networks for cooperative multi-agent learning}},
  author={Sunehag, Peter and Lever, Guy and Gruslys, Audrunas and Czarnecki, Wojciech Marian and Zambaldi, Vinicius and Jaderberg, Max and Lanctot, Marc and Sonnerat, Nicolas and Leibo, Joel Z and Tuyls, Karl and others},
  journal={arXiv preprint arXiv:1706.05296},
  year={2017}
}

@article{rashid2020monotonic,
  title={{Monotonic value function factorisation for deep multi-agent reinforcement learning}},
  author={Rashid, Tabish and Samvelyan, Mikayel and De Witt, Christian Schroeder and Farquhar, Gregory and Foerster, Jakob and Whiteson, Shimon},
  journal={Journal of Machine Learning Research},
  volume={21},
  number={178},
  pages={1--51},
  year={2020}
}

@ARTICLE{10183362,
  author={Azim, Ali Waqar and Bazzi, Ahmad and Fatima, Mahrukh and Shubair, Raed and Chafii, Marwa},
  journal={IEEE Trans. on Vehicular Tech.}, 
  title={{Dual-Mode Time Domain Multiplexed Chirp Spread Spectrum}}, 
  year={2023},
  volume={72},
  number={12},
  pages={16086-16097},
  }

@article{de2020independent,
  title={{Is independent learning all you need in the starcraft multi-agent challenge?}},
  author={De Witt, Christian Schroeder and others},
  journal={arXiv preprint arXiv:2011.09533},
  year={2020}
}

@book{albrecht2024multi,
  title={{Multi-agent reinforcement learning: Foundations and modern approaches}},
  author={Albrecht, Stefano V and Christianos, Filippos and Sch{\"a}fer, Lukas},
  year={2024},
  publisher={MIT Press}
}

@ARTICLE{abbasi2024lora,
  author={Aldhaheri, Lameya and Alshehhi, Noor and Manzil, Irfana Ilyas Jameela and Khalil, Ruhul Amin and Javaid, Shumaila and Saeed, Nasir and Alouini, Mohamed-Slim},
  journal={IEEE Internet of Things J.}, 
  title={{LoRa Communication for Agriculture 4.0: Opportunities, Challenges, and Future Directions}}, 
  year={2025},
  volume={12},
  number={2},
  pages={1380-1407},
  doi={10.1109/JIOT.2024.3486369}}

@article{alvarez2022uplink, 
title={{Uplink Transmission Policies for LoRa-Based Direct-to-Satellite IoT}}, 
author={Alvarez, G. and Fraire, J.A. and Hassan, K.A. and Cespedes, S.}, 
journal={IEEE Internet of Things J.}, 
volume={9}, 
number={18}, 
pages={17058--17071}, 
year={2022}, 
doi={10.1109/JIOT.2022.3175285}}

@ARTICLE{tondo2024multiple,
  author={Tondo, Felipe Augusto and others},
  journal={IEEE Access}, 
  title={{Multiple Channel LoRa-to-LEO Scheduling for Direct-to-Satellite IoT}}, 
  year={2024},
  volume={12},
  number={},
  pages={30627-30637}}

@ARTICLE{11165344,
  author={Lin, Kaiqiang and Alouini, Mohamed-Slim},
  journal={IEEE Internet of Things J.}, 
  title={{Connectivity Analysis of LoRaWAN-Based Nonterrestrial Networks for Subterranean mMTC}}, 
  year={2025},
  volume={12},
  number={23},
  pages={50604-50616},
  doi={10.1109/JIOT.2025.3610298}}

@ARTICLE{9537682,
  author={Ullah, Muhammad Asad and Mikhaylov, Konstantin and Alves, Hirley},
  journal={IEEE Trans. on Industrial Informatics}, 
  title={{Enabling mMTC in Remote Areas: LoRaWAN and LEO Satellite Integration for Offshore Wind Farm Monitoring}}, 
  year={2022},
  volume={18},
  number={6},
  pages={3744-3753},
  doi={10.1109/TII.2021.3112386}}

@ARTICLE{marchese2019flying,
  author={Marchese, Mario and Moheddine, Aya and Patrone, Fabio},
  journal={2019 Int. Symposium on Advanced Electrical and Comm. Technologies}, 
  title={{Towards Increasing the LoRa Network Coverage: A Flying Gateway}}, 
  year={2019},
  volume={},
  number={},
  pages={1-4},
  doi={10.1109/ISAECT47714.2019.9069697}}

@article{XIONG2023109511,
author = {Runqun Xiong and Chuan Liang and Huajun Zhang and Xiangyu Xu and Junzhou Luo},
title = {{FlyingLoRa: Towards energy efficient data collection in UAV-assisted LoRa networks}},
journal = {Computer Networks},
volume = {220},
pages = {109511},
year = {2023},
issn = {1389-1286},
doi = {https://doi.org/10.1016/j.comnet.2022.109511},}

@article{amichi2019spreading,
  author={Amichi, Licia and Kaneko, Megumi and Rachkidy, Nancy El and Guitton, Alexandre},
  journal={IEEE International Conference on Commun.}, 
  title={{Spreading Factor Allocation Strategy for LoRa Networks Under Imperfect Orthogonality}}, 
  year={2019},
  volume={},
  number={},
  pages={1-7},
  doi={10.1109/ICC.2019.8761235}}

@article{hamdi2020dynamic,
  author={Hamdi, Rami and Qaraqe, Marwa and Althunibat, Saud},
  journal={IEEE International Conference on Commun.}, 
  title={Dynamic Spreading Factor Assignment in LoRa Wireless Networks}, 
  year={2020},
  volume={},
  number={},
  pages={1-5},
  doi={10.1109/ICC40277.2020.9149243}}

@ARTICLE{su2020energy,
  author={Su, Binbin and others},
  journal={IEEE Trans. on Commun.}, 
  title={{Energy Efficient Uplink Transmissions in LoRa Networks}}, 
  year={2020},
  volume={68},
  number={8},
  pages={4960-4972},
  doi={10.1109/TCOMM.2020.2993085}}

@article{arasu2022efficient,			
  title={{Efficient Heuristic for Optimal MILP-LoRa Adaptive Resource Allocation for Aquaculture.}},
  author={Arasu, M Iniyan and Rani, S Subha and Geoffery, G Raswin},			
  journal={Intelligent Automation \& Soft Computing},			
  volume={33},			
  number={2},			
  year={2022}			
}

@article{bezdek2003convergence,
  title={{Convergence of alternating optimization}},
  author={Bezdek, James C and Hathaway, Richard J},
  journal={Neural, Parallel \& Scientific Computations},
  volume={11},
  number={4},
  year={2003}
}

@ARTICLE{10628007,
  author={Wang, Lei and Liang, Hongbin and Tang, Yanmei and Mao, Guotao and Zhang, Han and Zhao, Dongmei},
  journal={IEEE Internet of Things J.}, 
  title={{DRL-Based Joint Resource Allocation and Platoon Control Optimization for UAV-Hosted Platoon Digital Twin}}, 
  year={2024},
  volume={11},
  number={22},
  pages={37114-37126},
  doi={10.1109/JIOT.2024.3439576}}

@INPROCEEDINGS{9860807,			
author={Azizi, Farzad and others},	
booktitle={IEEE 95th Vehicular Technology Conference}, 	
title={{MIX-MAB: Reinforcement Learning-based Resource Allocation Algorithm for LoRaWAN}}, 			
year={2022}}

@ARTICLE{10540417,				
author={Rao, M. Rajeswara and Sundar, S.},				
journal={IEEE Access}, 				
title={{Enhancement in Optimal Resource-Based Data Transmission Over LPWAN Using a Deep Adaptive Reinforcement Learning Model Aided by Novel Remora With Lotus Effect Optimization Algorithm}}, 	
year={2024}}

@article{aihara2019q,				
 title={{Q-learning aided resource allocation and environment recognition in LoRaWAN with CSMA/CA}},		
author={Aihara, Naoki and Adachi, Koichi and Takyu, Osamu and Ohta, Mai and Fujii, Takeo},
journal={IEEE Access},				
volume={7},				
year={2019}}

@inproceedings{jouhari2023deep,				
title={{Deep reinforcement learning-based energy efficiency optimization for flying LoRa gateways}},
author={Jouhari, Mohammed and Ibrahimi, Khalil and Othman, Jalel Ben and Amhoud, El Mehdi},			
booktitle={IEEE International Conference on Comm.},	
 year={2023}}

@article{ZHAO2023100776,
title = {{Optimizing energy efficiency of LoRaWAN-based wireless underground sensor networks: A multi-agent reinforcement learning approach}},
author = {Guozheng Zhao and Kaiqiang Lin and David Chapman and Nicole Metje and Tong Hao},
journal = {Internet of Things},
volume = {22},
pages = {100776},
year = {2023},
issn = {2542-6605},
doi = {https://doi.org/10.1016/j.iot.2023.100776}}

@article{10437219,
  author={Zhang, Xu and Lin, Ziqi and Gong, Shimin and Gu, Bo and Niyato, Dusit},
  journal={2023 IEEE Global Commun. Conference}, 
  title={{Multiagent Reinforcement Learning with an Attention Mechanism for Improving Energy Efficiency in LoRa Networks}}, 
  year={2023},
  volume={},
  number={},
  pages={4152-4157},
  doi={10.1109/GLOBECOM54140.2023.10437219}}

@article{yu2020multi,
  author={Yu, Yi and Mroueh, Lina and Li, Shuo and Terr{\'e}, Michel},
  journal={31st Annual International Symposium on Personal, Indoor and Mobile Radio Commun.}, 
  title={{Multi-Agent Q-Learning Algorithm for Dynamic Power and Rate Allocation in LoRa Networks}}, 
  year={2020},
  volume={},
  number={},
  pages={1-5},
  doi={10.1109/PIMRC48278.2020.9217291}}

@ARTICLE{waret2018lora,
  author={Waret, Antoine and Kaneko, Megumi and Guitton, Alexandre and El Rachkidy, Nancy},
  journal={IEEE Wireless Communications Letters}, 
  title={{LoRa Throughput Analysis With Imperfect Spreading Factor Orthogonality}}, 
  year={2019},
  volume={8},
  number={2},
  pages={408-411},
  }

@inproceedings{slabicki2018adaptive,
  author    = {Slabicki, Mariusz and Premsankar, Gopika and Di Francesco, Mario},
  title     = {{Adaptive Configuration of LoRa Networks for Dense IoT Deployments}},
  booktitle = {2018 IEEE/IFIP Network Operations and Management Symposium (NOMS)},
  year      = {2018}
}

@article{mustari2024cooperative,
  title={{Cooperative THz communication for UAVs in 6G and beyond}},
  author={Mustari, Nazifa and Karabulut, Muhammet Ali and Shah, AFM Shahen and Tureli, Ufuk},
  journal={Green energy and intelligent transportation},
  volume={3},
  number={1},
  pages={100131},
  year={2024},
  publisher={Elsevier}
}

@article{eskandari2025visual,
  title={{Visual GANs for End-to-End UAV Trajectory Generation in RIS-Assisted Energy-Efficient Wireless Vehicular Networks}},
  author={Eskandari, Mohsen and Savkin, Andrey and Deghat, Mohammad},
  journal={Green Energy and Intelligent Transportation},
  pages={100365},
  year={2025},
  publisher={Elsevier}
}

@inproceedings{petajajarvi2015coverage,
  author    = {Pet{\"a}j{\"a}j{\"a}rvi, Juha and Mikhaylov, Konstantin and Roivainen, Antti and H{\"a}nninen, Tuomo and Pettissalo, Marko},
  title     = {On the coverage of {LPWANs}: Range evaluation and channel attenuation model for {LoRa} technology},
  booktitle = {Proc. 14th Int. Conf. ITS Telecommunications (ITST)},
  year      = {2015}
}

@article{brockman2016openai,
  title={Openai gym},
  author={Brockman, Greg and Cheung, Vicki and Pettersson, Ludwig and Schneider, Jonas and Schulman, John and Tang, Jie and Zaremba, Wojciech},
  journal={arXiv preprint arXiv:1606.01540},
  year={2016}
}

@inproceedings{magrin2017performance,
  title={{Performance evaluation of LoRa networks in a smart city scenario}},
  author={Magrin, Davide and Centenaro, Marco and Vangelista, Lorenzo},
  booktitle={IEEE International Conference on communications},
  year={2017},
}

@inproceedings{fan2019hybrid,
author = {Fan, Zhou and Su, Rui and Zhang, Weinan and Yu, Yong},
title = {{Hybrid actor-critic reinforcement learning in parameterized action space}},
year = {2019},
isbn = {9780999241141},
publisher = {AAAI Press},
booktitle = {Proceedings of the 28th International Joint Conference on Artificial Intelligence},
pages = {2279–2285},
numpages = {7},
location = {Macao, China},
series = {IJCAI'19}
}
\end{document}